\documentclass[]{jfm}
\usepackage{graphicx}
\usepackage{caption}
\usepackage{newtxtext}
\usepackage{newtxmath}
\usepackage{natbib}
\usepackage{hyperref}
\usepackage{subcaption}
\usepackage{float}

\usepackage{tikz}
\usepackage{xcolor}

\hypersetup{
    colorlinks = true,
    urlcolor   = blue,
    citecolor  = black,
}

\newcommand{\RomanNumeralCaps}[1]
\linenumbers

\definecolor{mplblue}{RGB}{0,114,189}
\definecolor{mplred}{RGB}{213,94,0}
\definecolor{mplgreen}{RGB}{86,177,76}

\newcommand{\ccontline}[1]{$\vcenter{\hbox{{\protect\tikz{\protect\draw[#1,-,line width=1.0pt] (0,0) -- (0.4,0);}}}}$}
\newcommand{\cdashedline}[1]{$\vcenter{\hbox{{\protect\tikz{\protect\draw[#1,dashed,line width=1.0pt] (0,0) -- (0.4,0);}}}}$}

\title{Reynolds effects on transition to turbulence for hypersonic expansion and compression corner flows}

\author{Clément Caillaud\aff{1}
  \corresp{\email{clement.caillaud@cea.fr}},
  Mathieu Lugrin\aff{2},
  Nicolas Severac\aff{2}
 \and Sébastien Esquieu\aff{1}}

\affiliation{\aff{1} CEA, CESTA, 15 av. des sablières, Le Barp - France
\aff{2} DAAA, ONERA, Institut Polytechnique de Paris, F-92190 Meudon - France}

\begin{document}
\maketitle

\begin{abstract}
An experimental and numerical investigation of transition to turbulence in attached boundary layer and through separated shock-boundary layer interaction is performed for a Cone-Cylinder-Flare geometry in the cold hypersonic regime at a Mach number of 7 and for a wide range of Reynolds numbers. 
The experimental campaign is conducted in the R2Ch facility and permits the collection of unsteady wall pressure fluctuations and high-speed schlieren images for all flow regimes. 
The collected data is then post-processed using data-driven analysis and compared to base flow computations and global linear stability analysis to better understand the mechanisms at stake in the transition to turbulence observed in the experiments. Consistent with previous studies, the trends in Reynolds from the experimental data show a strong variation of the length of the separated region depending on the upstream state of the boundary-layer. The results enable to distinguish two kinds of transition regimes which were not clearly defined before.
For the high Reynolds number cases, transition is found to be dominated by second Mack mode and its non-linearities on the cone. High-frequency wall pressure measurements and schlieren imaging of both the fundamental wave and the non-linear harmonics are provided. Non-linear interaction regions are observed with unprecedented resolution, helping to understand the state of the boundary-layer before its rapid breakdown at the reattachment point.
At lower Reynolds numbers, the transition scenario is more complex with the coexistence of both low- and high-frequency modes. A complex coupling between the separated flow and the dominating convective instabilities is highlighted. Trapped acoustic waves inside the recirculation region are clearly measured for the first time to the best of the authors' knowledge. Their linear origin is demonstrated using global linear stability analysis and a simple acoustic duct model is provided to predict their frequencies. These waves provide a new perspective on the low-frequency first-mode pressure signature at the wall and another interpretation is provided. Finally, the role of these acoustic modes in an energy transfer from high to low frequencies, leading to transition on the flare, is assessed.
\end{abstract}
    
\section{Introduction}
    Whether they are cruising or re-entering the atmosphere, it is commonly accepted that hypersonic vehicles face strong aerothermal loads on their external surfaces. In the peculiar case of reentry flight, the boundary layer on the vehicle generates substantial friction, heat, and mechanical loads. If not accounted for, the combined action of these solicitations can be so intense that they may directly alter the integrity of the vehicle's Thermal Protection System (TPS) or its controllability, which could both lead to its loss. Therefore, from an engineering perspective, being able to understand and maybe predict the occurrence of such critical heating conditions is of paramount importance in the sizing of the TPS and the internal structure of a reentry vehicle. Especially considering that in the system design phase, an overestimation of such aerothermal loads may lead to a substantial loss of performance due to an excess of TPS mass, while an underestimation may produce a vehicle unable to survive the actual reentry loads. 
    \begin{figure}
        \def\svgwidth{\textwidth}
        \input{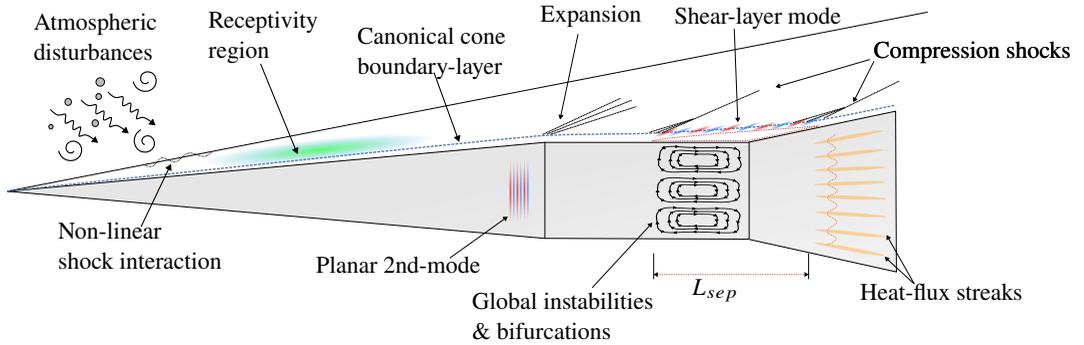}
        \caption{Summary of the Cone-Cylinder-Flare hypersonic-flow features}
        \label{fig:summary-ccf}
    \end{figure}

    One of the key physical phenomena at play in the generation of these aerothermal loads is the boundary-layer transition to turbulence. Research in high-speed aerodynamics has shown over the past 60 years that the transition in the hypersonic regime can locally produce up to a tenfold increase in the wall heat flux compared to a laminar boundary layer. Additionally, there are drastic changes in viscous efforts and pressure distributions that directly impact the flight quality of the vehicle, justifying important efforts in understanding and modeling the effect of transition \citep{andersonjr.HypersonicHighTemperatureGas2006,schneiderHypersonicLaminarTurbulent2004}. The route to turbulence can follow various paths depending on the geometry and flow conditions and remains very sensitive to small variations in these parameters. Notably, the shape of a vehicle can induce various flow topologies : canonical boundary layers, crossflow effects, separations and reattachments, centreline vortices, favourable and adverse pressure gradients, entropy-layers, wakes, etc. Many previous studies have attempted to investigate such vehicle geometry effects. We can list the HiFire program \citep{dolvin2008hypersonic,juliano2015hifire}, BOLT \citep{wheaton2018boundary} or more recently the Cone-Cylinder-Flare geometry \citep{esquieuFlowStabilityAnalysis2019}. The latter is the object of this combined experimental and numerical study. Despite its simple configuration, the CCF geometry, depicted in figure~\ref{fig:summary-ccf} offers many advantages as it can simultaneously allow for the study of simple developing laminar boundary-layers, bluntness effects, transitional and turbulent structure interaction with expansion fans and most importantly separation induced by shock-boundary layer interactions. Moreover, at zero angle of attack, its cylindrical frame enables performing advanced numerical studies such as global stability analysis or DNS, and it also eases its optical observation in an experimental setup.

\begin{figure}
        \centering
        \includegraphics[width=0.9\textwidth]{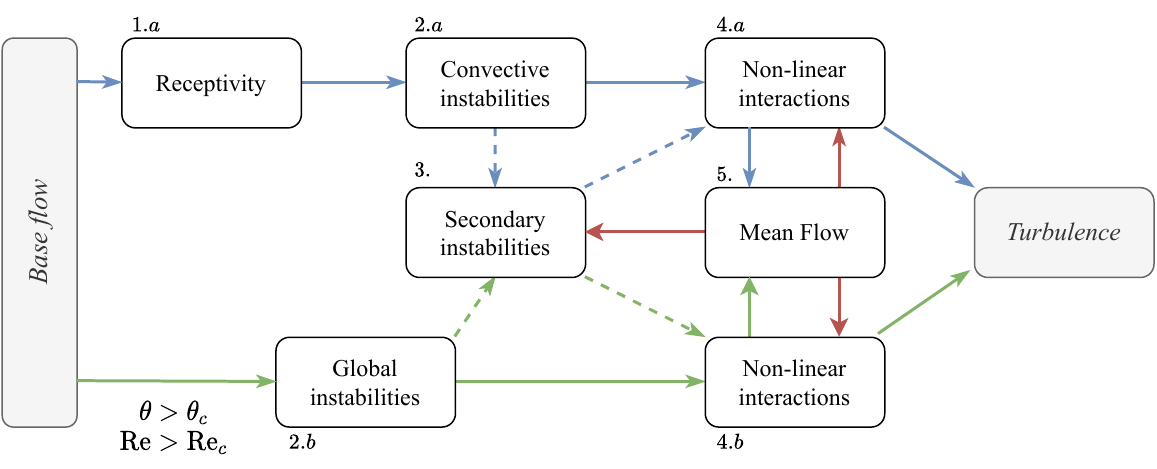}
        \caption{Overview of the possible routes to turbulence for a CCF-like geometry}
        \label{fig:routes-turbulence}
    \end{figure}

    The CCF boundary-layer transition can follow different paths depending on the receptivity, the Reynolds number and the flare angle \cite{paredes2022boundary}. These paths are summarized in figure~\ref{fig:routes-turbulence}. Starting from a steady laminar baseflow , the turbulence can theoretically originate from either the convective or the global, modal, instabilities, (see routes $a$ or $b$). These instabilities relate to the noise-amplifier or oscillator nature of the flow, respectively \citep{huerreAbsoluteConvectiveInstabilities1985}. For the convective routes, previous experimental and numerical studies, \citep{benitezInstabilityMeasurementsAxisymmetric2020, benitezInstabilityTransitionOnset2023, paredes2022boundary, caillaud2025separation} have shown that three families of instabilities may exist on the aft sharp cone in cold isothermal wall conditions. Namely, the first- and second- Mack's modes and streamwise streaks. The former two are the classical modal instabilities of the hypersonic boundary layer \citep{mackBoundaryLayerLinearStability1984a} while the latter comes from the non-modal lift-up mechanism \citep{landahlNoteAlgebraicInstability1980}. These three instabilities are seeded by the disturbance environment, such as acoustic, vorticity and entropy waves from the shock/free-stream interaction or even small particulates that penetrates below the shock (step 1.a). On the cone region, previous studies found the second-mode instability to be the most amplified for cold-wall conditions. 
    Downstream of the cone, these instabilities first interact with the expansion at the cone-cylinder junction and then enter the separated flow-region which is characterised by a viscous interaction with the compression shock created by the flare. Depending on the  flare angle and Reynolds number, the adverse pressure gradient caused by the shock may be sufficient to cause boundary layer separation, which creates a recirculation region at the base of the flare. 

    In the studied configuration, the shock-wave laminar boundary-layer interaction (SWBLI) in this region produces a complex separated flow topology, which for hypersonic flows, is strongly coupled with the transitional dynamics of the boundary-layer \citep{marxenEffectSmallamplitudeConvective2011,lugrinTransitionScenarioHypersonic2021}. 
    The complexity of this region is further heightened by the presence of global modal instabilities in the separated region. As illustrated in figure~\ref{fig:routes-turbulence} with the step 2.b, for a critical flare-angle $\theta_c$ and/or critical Reynolds number $\Rey_c$, global instabilities may lead the flow to bifurcate to 3D steady or unsteady states \citep{robinet2007bifurcations,paredes2022boundary} as illustrated in figure~\ref{fig:summary-ccf}. These global modes are located either within the recirculation bubble or at the boundary-layer reattachment on the flare \citep{dwivediReattachmentStreaksHypersonic2019, caoTransitionTurbulenceHypersonic2022}.
    At the same time, the shear layer above the separated region acts as a strong amplifier for some incoming convective instabilities. It should be noted that these convective instabilities are most often dominating the measurements, which actually makes the experimental observation of global instabilities difficult, even questioning their relevance for transition in conventional hypersonic wind-tunnel environment \citep{lugrinMultiscaleStudyTransitional2022}. Oblique waves, having similar frequencies to the first-mode are observed to destabilize within the shear layer \citep{dwivediObliqueTransitionHypersonic2022,lugrinTransitionScenarioHypersonic2021,caoUnsteadyEffectsHypersonic2021, haoResponseHypersonicCompression2023}. The actual amplification mechanism of these oblique waves remains unclear and it has been hypothetised to be driven by different independent or combined mechanisms such as: first-mode instability, shear layer instability akin to the Kelvin-Helmotz mechanism, or even unsteady Görtler instabilities in the curved regions at separation and reattachment \citep{Song_Hao_2025}. On the other hand, second mode waves are found to be weakly amplified, or even stabilized in the separation region. However, these waves get destabilised as soon as the boundary-layer reattaches. This evolution of the second-mode in the separated region also remains unclear and recent research have discussed a possible role of higher-order Mack modes in the recirculation bubble \citep{caillaud2025separation}.
    Finally, steady longitudinal streaks are known to get substantially amplified and lead to an intense wall heating at the boundary layer reattachment (figure~\ref{fig:summary-ccf}).
    These streaks were observed experimentally on similar ramp configurations \citep{lugrin2022transitional,benitezInstabilityTransitionOnset2023,benitezMeasurementsBluntConeCylinderFlarea}, their origin may vary depending on the Reynolds and Mach numbers and how amplified are the incoming boundary layer instabilities in the separated region.
    Different experiments and numerical studies identified the origin of these streaks as: the lift-up mechanism in the incoming boundary-layer; the byproduct of the non-linear interaction of two opposed oblique convective instabilities; the non-linear saturation of global instabilities; steady Görtler instabilities along the curved streamlines of the separated flow; baroclinic torque lifting up the low-momentum flow from the wall.
    
    Reverting to figure~\ref{fig:routes-turbulence}, the aforementioned global and convective instabilities along with their eventual secondary-instability (step 3.) all contribute to non-linear interactions (step 4. a and b) through triad-waves relations or mean flow distortions \citep{craikNonlinearResonantInstability1971,craigNonlinearBehaviourMack2019,kuehlNonlinearSaturationNonlinear2017}.
    In the next stages, they produce a complex interplay within the viscous-interaction region where the separated flow directly evolves with the incoming convective instabilities from the cone region. When instabilities wavepackets bring enough energy in the separated boundary-layer to cause breakdown, it results in a shrinking of separated region and thus a reduction of the separation length $L_{sep}$ linked to the amplitude of the incoming disturbances from the cone region and the ability of the mixing layer to amplify those disturbances \citep{lugrinTransitionScenarioHypersonic2021}. Therefore, for given Reynolds number and transitional dynamics (which is dependent on noise environment, wall temperature, etc) the separated region can be observed to have very different topologies, as illustrated recently by \cite{benitez2025separation} on a blunt CCF geometry in two hypersonic wind-tunnels. 

    Considering the numerous possibilities in routes to transition offered by the CCF geometry. The present research aims at clarifying experimentally observed transition dynamics as the Reynolds number and the separation length vary. We seek to combine experiments and numerical analysis to understand what are the possible transition scenarios for different Reynolds regimes. The emphasis being put on the clear identification of the instabilities properties and their interactions as the Reynolds number increases and the separation region interacts with the incoming boundary-layer disturbances. The analysis is outlined as follows. First, the experimental setup and facilities are detailed (\ref{sec:expe-setup}), then the theoretical framework of global stability analysis and associated experimental post-treatments methods are provided (\ref{sec:num-tools} and \ref{sec:expe-tools}). The results section starts with an overview of the Reynolds effects on the experimental mean flow and the numerical baseflows, followed by an overview of the sensors response (\ref{sec:expe-overview}). From this summary, two observed dominant transition scenarios are discussed, namely a short (\ref{sec:small-sep}) and a long separation (\ref{sec:large-sep}) routes. Finally, concluding remarks on the findings of this numerical and experimental investigation are provided (\ref{sec:discussion}).



\section{Experimental Setup} \label{sec:expe-setup}
\begin{figure}
    \begin{subfigure}[b]{0.49\textwidth}
        \centering
        \includegraphics[width=\textwidth]{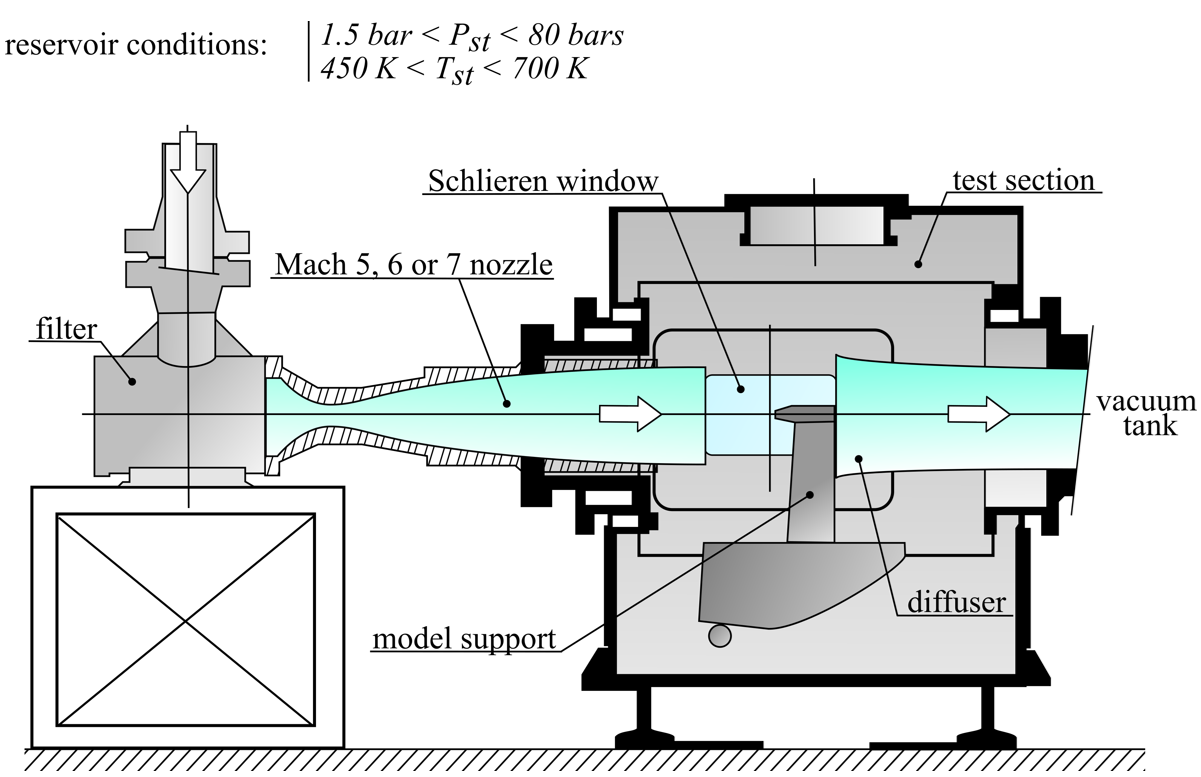}
        \caption{R2Ch blow-down wind-tunnel}
        \label{fig:R2Ch}
    \end{subfigure}
    \hfill
    \begin{subfigure}[b]{0.49\textwidth}
        \centering
        \includegraphics[width=\textwidth, angle=0]{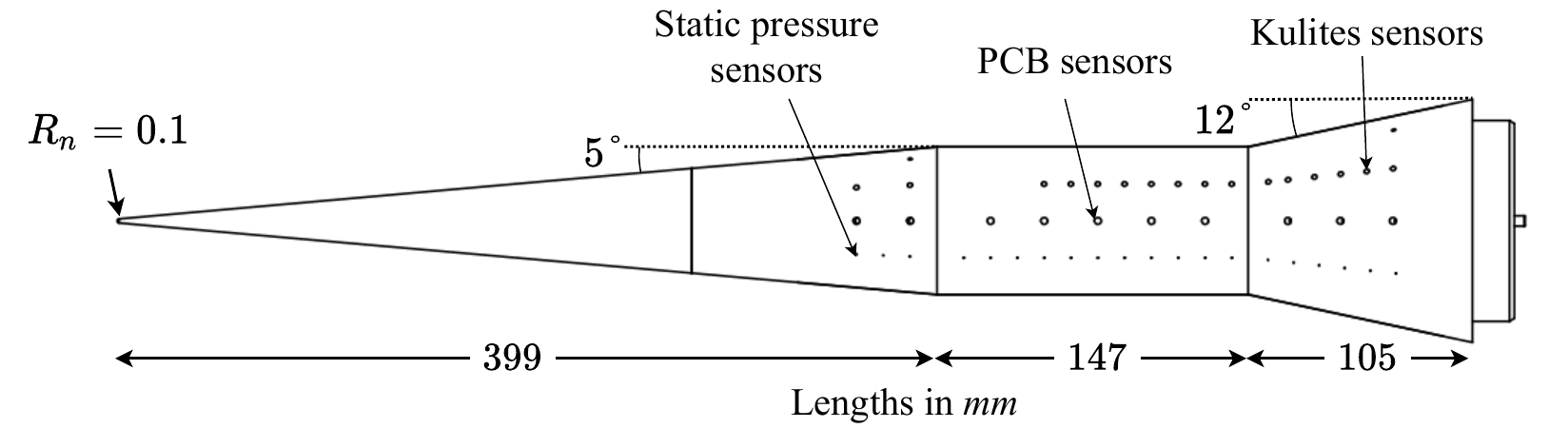}
        \caption{CCF12 model used in the experiments}
        \label{fig:ccf12-model}        
    \end{subfigure}
\caption{Overview of the wind-tunnel and the CCF12 model used for the campaign.}
\label{fig:R2Ch_and_ccf12}
\end{figure}


The experiments are carried out in France at ONERA Meudon, using the R2Ch low enthalpy hypersonic blow-down wind tunnel depicted in figure~\ref{fig:R2Ch}. This facility allows for exploring flows with Mach numbers of 5, 6, and 7 using a set of three exchangeable contoured nozzles. Although runs were conducted at Mach numbers of 6 and 7, only results at a Mach number of 7 will be presented in this article. The stagnation conditions of the facility are in the range of $0.5\leq P_0 \leq 80$~bars and $450\leq T_0 \leq 750$K which permits exploring an extended range of Reynolds numbers in the cold hypersonic flow regime. The outlet of the nozzle has a diameter of $327$~mm and opens to an open-jet test section in which the model is placed as shown in figure \ref{fig:R2Ch}. A diffuser placed downstream of the model collects the hypersonic flow and directs it to a $500\, \text{m}^3$ vacuum sphere. The total stabilised test time extends to the minute. The wind-tunnel noise is considered as conventional and static pressure fluctuations in the free stream has been measured to be less than $1.5\%$ and at rather high frequencies \citep{threadgill2025scaling}.

The CCF12 model studied in the R2Ch test section is shown in figure \ref{fig:ccf12-model}. It is a standard polished steel model ($Ra<1\,\upmu\text{m}$), consisting of an interchangeable nosetip. The $0.1$\,mm nose radius is used for the present study (the step at the junction between the nose and the body of the model, located $273$\,mm downstream of the nosetip, has been measured to be less than $25 \, \upmu \text{m}$ and is thus negligible compared to the boundary layer thickness), followed by a $5^\circ$ half angle, $399.3$\,mm long cone, a $147$\,mm long cylinder, and finally a $12^\circ$ half angle flare that extends the length of the full model up to $650$\,mm.
All results are presented with an origin located at the cone-cylinder junction of the sharp model. Thus, cone data is set at negative $x$ values and cylinder-flare at positive $x$ values.
The model is mounted on a static mount and set at a $0^\circ$ angle of attack (AOA) using four diametrically opposed static pressure taps at the end of the cone to estimate the angle of attack until the pressures are all equal (up to the sensor precision, which is around 30\,Pa, leading to an uncertainty in the true angle of attack and yaw of less than $0.1^\circ$).

In order to probe the flow dynamics, the measurements use two kinds of wall-pressure sensors to capture steady and high-frequency dynamics. Additionally, optical measurements are performed using high-speed Z-type schlieren for different fields of view (FOV) around the object. These diagnostics are further detailed in the following sections. 

\subsection{Wall-pressure sensors}
\begin{table}
    \centering
    \begin{tabular}{l l c l c}
        
        Regime            & Name        & Number & Sampling Rate & Range \\ \hline
        Quasi-steady      & Static Taps & 20     & 100 Hz        & N/A \\
        High-frequency    & PCB 132B38  & 10     & 3.3 MHz       & [10, 450] kHz \\
        High-frequency (1 run)    & IC2 MHz  & 1     & 2 MHz       & [1, 800] kHz \\
        \hline
    \end{tabular}
    \caption{Wall pressure sensors summary}
    \label{tab:wp-sensors}
\end{table}
The wall pressure measurements are performed using  azimuthally separated streamwise arrays of sensors along the CCF12 geometry as depicted in figure \ref{fig:ccf12-model}, these sensor details are provided in Tab.~\ref{tab:wp-sensors}. The axial position of the sensors is detailed in  Tab.~\ref{tab:wp-sensors-pos}.

\begin{table}
\resizebox{\textwidth}{!}{
    \centering
    \begin{tabular}{lllllllllllllllllllllllll}
    \hline
    Sensor number & 01 & 02 & 03 & 04 & 05 & 06 & 07 & 08 & 09 & 10 & 11 & 12 & 13 & 14 & 15 & 16 & 17 & 18 & 19 & 20 \\
    \hline
    PCB & -38.3 & -13.3 & 24.7 & 50.7 & 75.7 & 101.7 & 126.7 & 165.7 & 190.7 & 215.7 & - & - & - & - & - & - & - & - & - & - \\
    \hline
    Static & -38.3 & -25.3 & -13.3 & 12.7 & 24.7 & 37.7 & 50.7 & 63.7 & 75.7 & 88.7 & 101.7 & 113.7 & 126.7 & 139.7 & 156.7 & 168.7 & 180.7 & 192.7 & 204.7 & 216.7 \\
    \hline
    \end{tabular}}

        \caption{Wall pressure sensors position measured in mm from the cone-cylinder junction}
    \label{tab:wp-sensors-pos}
\end{table}

The static pressure taps are connected by means of vinyl tubes to TE 16MS microscanners placed close (around 1\,m) to the model to reduce the response time of the vinyl. The wall pressure is sampled at 100\,Hz giving an estimation of pressure profile in order to assess the impact of the Reynolds number on key flow features, such as separation and reattachment points. The 16Ms are connected to Chell CANDAQ5 and microDAQ-Int for digitization and communication with the acquisition system.


High-frequency wall measurements are made with PCB 132B38 sensors. 
To ensure flush integration of the sensors on the wall, an epoxy cap is applied at the top of the transducer and then surfaced to match the curvature profile of the model following the procedure described in \cite{lugrin2022transitional}.
The PCB 132B38 signal is sampled at 3.3\,MHz by a National Instrument PXI-6376 card after conditioning by a PCB 482C05, the signal is neither amplified nor filtered. 
In the current setup, and given the electronic noise environment of R2Ch, the PCB sensors are found to provide accurate measurements in the $[11,500]$~kHz frequency range, Noise floors are always presented throughout the study to avoid misinterpreting the signal coming from non-aerodynamic sources.
For a limited amount of runs, the most downstream PCB sensor on the cone of the model (PCB02) was replaced by a IC2 MHz sensor, which was conditioned and sampled at 2\,MHz by a Dewetron Multi-1820 card.

\subsection{High-speed schlieren}

High-speed schlieren measurements have been conducted during the campaign. They complement the local wall pressure measurements by providing time-resolved measurements of the vertical density gradients for different fields of view around the geometry. The setup consists of a standard Z-type schlieren setup using a Phantom TMX7510 camera illuminated by a continuous and punctual source using a $300$~W xenon lamp. 
The camera frequency used in this study is $850$~KHz with an exposure time of $0.2 \mu s$ and a spatial resolution of $640 \times 128$ pixels in binned mode. 
The high sampling frequency of the density gradient permits obtaining time resolved datasets of the flow dynamics and uses data decomposition in the Fourier space as introduced in the next section.  
In order to increase the convergence of spectral estimators, the full camera memory capabilities (128Go) are used, leading to more than 2 million frames per run.  

        


\section{Methods for numerical operator-based Analyses} \label{sec:num-tools}
Probing the flow physics using the aforementioned experimental setup and the associated analyses (PSD, SPOD, BMD) enables highlighting the most energetic features of the transition process. However, identifying the actual nature of the structures that are growing in the flow can be cumbersome as a result of a partial picture of the physics being captured. Hence, an operator-based approach using the linearised Navier-Stokes equations and the associated Resolvent operator allows the formal identification of absolute and convective global instabilities growing around the laminar baseflow in a comprehensive manner. This linear stability framework is summarised in the next sections.

\subsection{Governing equations} \label{sec:BFC}
A first step in the study of the CCF transition dynamics is the computation of steady laminar baseflows that support the transition instabilities. Particularly, by considering low Reynolds number and weak non-linear effects, the computed steady solutions will be shown to remain close to the experimental mean-flows. The steady solutions are computed with the state vector $\pmb q(\pmb x, t)$ describing the time evolution of the conservative variables vector $(\rho, \rho u_x, \rho u_r, \rho u_\theta, \rho E)^T$  in the cylindrical coordinates $\pmb x = (x, y, \theta)$, the discrete non-linear Navier-Stokes equations can be written as the forced non-linear dynamical system,
\begin{align}
    \frac{\partial }{\partial t} \pmb{q}(\pmb x, t) = \mathsfbi{N}(\pmb q (\pmb x, t)) + \pmb f_e(\pmb x, t). \label{eq:NS-DS}
\end{align}

With $\mathsfbi N$ the discrete non-linear operator of the compressible Navier-Stokes equations and $\pmb f_e$ the external sources of disturbances. The operator $\mathsfbi N$ uses the perfect gas model : $p = \rho R_{air} T$ and the Sutherland law for viscosity.

\subsection{Resolvent Analysis} \label{sec:GSA}
The linear instabilities leading to turbulence around the CCF geometry are identified using linear stability theory commonly used in transition to turbulence studies \citep{schmidStabilityTransitionShear2012}. This mathematical framework permits to investigate the initial amplification of small amplitude disturbances in the boundary-layer. In this section the key concepts used to characterize the forced response of the CCF flow in the experiment are summarised. 

Starting from, \ref{eq:NS-DS}, the linear stability analysis of the Navier-Stokes dynamical system is performed by considering the following state decomposition, $\pmb q(\pmb x,t) = \pmb q_0(\pmb x) + \varepsilon \pmb q'(\pmb x, t),\  \varepsilon \ll 1$. With $\pmb q_0(\pmb x)$, a steady state, termed the baseflow, such that $\pmb{\mathcal{N}}(\pmb q_0)=0$ and $\pmb q'(\pmb x, t)$ a very small disturbance component of order $\epsilon$, representing fluctuations around $\pmb q_0$. 

Using this decomposition of the state and performing a series expansion to the first order of equation~\ref{eq:NS-DS}, the linearised Navier-Stokes equations can be rewritten as :
\begin{equation}
        \frac{\partial}{\partial t} \pmb q'(\pmb x, t) = \mathsfbi{J}\pmb q' (\pmb x, t)  + \pmb f_e(\pmb x, t). \label{eq:NS-DS-LIN}
\end{equation}
With $\mathsfbi{J}$ the Jacobian matrix of the discrete Navier-Stokes equations linearised around the base-flow $\pmb q_0$.
From this linear dynamical system, one can either study the self-sustained dynamics of the flow when $\pmb f_e(\pmb x, t)=0$ or the noise amplifier dynamics of the forced flow when $\pmb f_e(\pmb x, t) \neq 0$, \citep{huerreLocalGlobalInstabilities1990}.


In the present study, with the exogenous forcing of the wind-tunnel, the flow mainly acts as noise-amplifier system. An efficient method to obtain the optimal flow structures driving the transition dynamics is the Resolvent operator analysis. This operator is obtained here by considering statistically stationary disturbances in the wind tunnel and axisymmetric flow conditions. Harmonic perturbations of the form $\pmb{f_e}=\pmb{\hat{f}}(x,y)e^{i\omega t+i m\theta}$ and $\pmb{q}'=\pmb{\hat{q}}(x,y)e^{i(\omega t + m\theta)}$ are used and equation~\ref{eq:NS-DS-LIN} can be written in the temporal domain as,
\begin{equation}
    \frac{\partial}{\partial t}  \pmb{\hat{q}}(x,y)e^{i\omega t+i m\theta} = \mathsfbi{J}\pmb{\hat{q}}(x,y)e^{i\omega t+i m\theta}  + \pmb{\hat{f}}(x,y)e^{i\omega t+i m\theta}. \label{eq:ns-lin-temporal}
\end{equation}
It should be noted that for this 2D axisymmetric formulation, the Jacobian matrix is defined for a given azimuthal wavenumber $m$, leading to $\mathsfbi{J} = \mathsfbi{J}(\pmb q_0, m)$ \citep{bugeat3DGlobalOptimal2019}. The Resolvent operator is obtained by writing the frequency domain counterpart of equation~\ref{eq:NS-DS-LIN} and factoring the terms to obtain an input-output relation between the linear flow response $\pmb{\hat{q}}$ and the incoming disturbance forces $\pmb{\hat{f}}$
\begin{equation}
    \pmb{\hat{q}} = (i\omega \mathsfbi{I}-\mathsfbi{J})^{-1}  \pmb{\hat{f}} \label{eq:input-output}
\end{equation}
The linear transfer function $\pmb{\mathcal{R}}=(i\omega \mathsfbi{I}-\mathsfbi{J})^{-1}$, with $\mathsfbi{I}$ the identity, is the Resolvent matrix of the system for a given baseflow $\pmb q_0$, frequency $\omega$ and wavenumber $m$. This operator allows the observation of the non-modal instability of the baseflow resulting from the non-normality of $\mathsfbi{J}$ \citep{trefethenHydrodynamicStabilityEigenvalues1993, schmidNonmodalStabilityTheory2007}. 

This representation of the linear dynamics of the Navier-Stokes equations through the Resolvent operator is particularly useful for the description of modal and non-modal convective instabilities in transition. Its Singular Value Decomposition (SVD), 
$\pmb{\mathcal{R}} = \mathsfbi{U}\pmb{\Sigma}\mathsfbi{V}$
provides for each frequency $\omega$ and wavenumber $m$, the matrix $\mathsfbi{U}$ of all the responses $\pmb\psi_i(x,y, t)$ and the matrix $\mathsfbi{V}$ of the forcings $\pmb\phi_i(\pmb x, t)$ supported by the system. The vectors in the forcings and responses orthogonal bases are ranked by decreasing importance as defined by the gain matrix $\pmb{\Sigma} = \text{diag} (\mu_0^2, \mu_1^2,...,\mu_N^2)$. For a given energy norm $\Vert.\Vert_E$, here chosen as the Chu norm \citep{chuEnergyTransferSmall1965}, the gain $\mu_i$, describes the energy ratio of the response and the forcing,
\begin{align}
    \mu_i^2 = \frac{\Vert\pmb \psi_i\Vert_E}{\Vert\pmb \phi_i\Vert_E}, \quad \text{with} \Vert . \Vert_E = \pmb [.]^T \mathsfbi W_E \pmb [.].
\end{align}
In this representation of the system, the special case of a large separation between the optimal gain and the first sub-optimal gain, $\mu_0^2 >> \mu_1^2$ means that a single mechanism is dominating the linear dynamics at the considered $(\omega, m)$ pair. Using the SVD of the Resolvent matrix, the response $\pmb q'$ of the flow to external disturbances $\pmb f_e$, given in Eqs~\ref{eq:input-output} and can be expressed as,
\begin{align}
    \pmb{\hat{q}} = \sum_{i=0}^{N} \pmb \psi_i \underbrace{\mu_i^2 \langle \pmb\phi_i, \pmb{\hat{f}} \rangle_E}_{c_0}. \label{eq:receptivity}
\end{align}

In this form, the observed linear responses $\pmb q'$ can be seen as the linear combination of optimal and sub-optimal responses $\pmb \psi_i$ weighed by the receptivity coefficient $c_0$ which accounts for the gain $\mu_i^2$, defining the efficiency of the linear mechanisms and the capacity of the external forcing $f_e$ to project energy on the optimal forcing basis $\pmb \phi_i$ at the given frequency. Further details about this framework can be found in \cite{sippDynamicsControlGlobal2010,beneddineConditionsValidityMean2016}. This representation of the linear dynamics will be extensively used in the results sections to identify and discuss the experimentally observed flow structures.

%
%


\subsection{Numerical setup and discretisation}
The equation~\ref{eq:NS-DS} is solved using the BROADCAST \citep{poulainBROADCASTHighorderCompressible2023} toolbox. 
The convective fluxes are computed using a $7^{th}$ order Flux-Extrapolated MUSCL scheme and the shock-capturing technique described in \cite{sciacovelliAssessmentHighorderShockcapturing2021}. The viscous fluxes are solved with a fourth-order accurate scheme on a five-point compact stencil. The baseflow, which is a fixed point of equation~\ref{eq:NS-DS}, is computed using a pseudo-transient method which tends to a Newton iteration scheme as the pseudo-timestep $\Delta t$ increases close to convergence  \citep{crivelliniImplicitMatrixfreeDiscontinuous2011},
\begin{equation} 
    \left( \frac{\pmb I}{\Delta t} + \left.\frac{\partial \mathsfbi{N}}{\partial q}\right|_{q} \right) \delta {\pmb q} = - \mathsfbi{N}(\pmb q (\pmb x)).
    \label{eq:Newton_steps} 
\end{equation}
This converged laminar baseflow is a prerequisite to the global linear stability described hereafter. The BROADCAST toolbox allows computing derivatives of the discrete Navier-Stokes equations with respect to a state $\pmb q_0$ using algorithmic differentiation. This approach provides exact discrete direct and adjoint linear operators. The inversion of the linear systems considered is then conducted using LU decomposition through PETSc, using the multifrontal MUMPS solver. This combination of methods makes possible the computation of fixed points of the system (\ref{eq:NS-DS}) up to machine precision.

The numerical setup, including detailed boundary conditions, mesh and mesh convergence study can be found in \cite{caillaud2025separation}. To reproduce the experimental setup, all cases are simulated with an isothermal wall at 300K.

\section{Methodology for experimental Data--Based Analyses} \label{sec:expe-tools}

In order to identify and characterize the flow dynamics observed through
high-speed schlieren, two mathematical frameworks are introduced. The Spectral Proper Orthogonal Decomposition (SPOD) \citep{lumleyStochasticToolsTurbulence1970} consists of a data-driven approach which allows educing spatio-temporally coherent structures from the high-speed schlieren data-sets. This method is further extended by the Bispectral Mode Decomposition (BMD) introduced by \cite{schmidtBispectralModeDecomposition2020}. The BMD allows us to uncover the possible triadic interactions between spatio-temporal coherent structures and is therefore a useful tool for understanding the initial non-linear stages of transition to turbulence.

\subsection{Spectral Proper Orthogonal Decomposition (SPOD)}\label{sec:spod}

SPOD is becoming a common tool to post-process high speed schlieren data of hypersonic instabilities in the last few years \citep{butlerInteractionSecondmodeWave2021,butlerTransitionalHypersonicFlow2022,lugrin2022transitional,benitez2025separation} as it allows the extraction of spatio-temporally coherent structures from the images. 
It has been first introduced by \cite{lumleyStochasticToolsTurbulence1970}. 
The procedure used here to compute the SPOD modes is close to the one described in 
\cite{schmidtGuideSpectralProper2020}.

First, frames $\pmb s \in \mathbb{R}^{N_p}$, stacked in time dependent vectors $\mathbf{b}(t) \in \mathbb{R}^{N_p\times N_t}$ corresponding to the image grey intensity levels, with $N_p$ the number of pixels and $N_t$ the number of frames per blocks, are stacked in $N_b$ non-overlapping blocks,
\begin{align}
    \mathsfbi{S}_t=\left(\pmb{b}_0(t),\pmb{b}_1(t), \dots \pmb{b}_{N_b}(t)\right), \quad \pmb{b}_i(t) = \pmb{b}(t_0 + i N_t)
\end{align}
containing a temporal sequence of length $N$ at discrete times $t$, going from the initial time $t_0$ to the final frame at $t_f$. The blocks are windowed using a Hanning window.
A Fourier transform in time $\pmb{\mathcal F}$ is performed on each block $\pmb{b}_i(t)$ to obtain vectors $\hat{\pmb b}^i$ in space at all resolved frequencies for each independent realisation,
\begin{align}
    \hat{\mathsfbi S}=\left(\hat{\pmb{b}}^{0}(f),\hat{\pmb{b}}^{1}(f),\dots \hat{\pmb{b}}^{N_b}(f) \right), 
    \quad f=\left\{f_0,...,f_k,...,f_{N_t/2 +1}\right\}.
\end{align}

At the core of the SPOD lies the selection of a set of vectors at a frequency $f_k$ from each independent block $\hat{\pmb b}^i$ to constitute a matrix $\hat{\mathsfbi X}_{k}$ of all the Fourier modes at $f_k$,
\begin{align}
    \hat{\mathsfbi{X}}_{k} =\left( \hat{\pmb{b}}^{0}(f_k), \  \hat{\pmb{b}}^{1}(f_k), \  \cdots, \  \hat{\pmb{b}}^{N_b}(f_k) \right), \quad  \hat{\mathsfbi{X}}_{k}\in\mathbb{C}^{N_p \times N_b}.
\end{align}
The cross spectral density matrix $ \hat{\mathsfbi{X}}_{k}\hat{\mathsfbi{X}}_{k}^H$ is then used to elicit spatially coherent structures between each realisation vector $\hat{\pmb{b}^i}$ at the frequency $f_k$. The singular value decomposition (SVD) of the cross spectral density matrix at $f_k$ is obtained through the snapshots method by solving the reduced eigenvalue problem,
\begin{align}
    \hat{\mathsfbi{X}}_{k}^H\hat{\mathsfbi{X}}_{k} \pmb{x} =\sigma\pmb{x}.
\end{align}
The decomposition allows us to recover an orthogonal basis of right singular vectors of $\hat{\mathsfbi{X}}_{k}$,  from which one can easily retrieve the left singular vectors, representing independent spatio-temporally coherent structures of density gradient at the frequency $f_k$. These vectors are ranked by importance (i.e. energy in the $L_2$ norm sense) by their singular values $\sigma_0 > \sigma_1 > \dots > \sigma_{N_b}$. It should be noted that formal links can be drawn between the SPOD of a flow field and the Resolvent optimal response basis of the associated baseflow \citep{towneSpectralProperOrthogonal2018}.

In practice, the size of the block $N_t$ is chosen to get a good compromise between the spectral resolution and the convergence of the SPOD, typically blocks of $N_t = 1024$ frames are chosen. Given the large amount of data collected, a number of blocks of $N_b=800$ is found to be sufficient. It should be noted that this combination of frames and blocks number is much more than the usual numbers used in hypersonic experimental studies. To allow an efficient computation of the SPOD modes for the different high-speed movies, the algorithm is implemented using distributed memory parallelisation to speed up the computation and distribute the data on a large cluster. This parallel computation allows post-processing entire 128Go movies at once to improve the convergence of the SPOD modes.

\subsection{Bispectral Mode Decomposition}\label{sec:bmd}
The Bispectral Mode Decomposition  extends the SPOD to higher-order spectra and provides insights on the quadratic interactions of the spatio-temporal coherent structures \citep{schmidtBispectralModeDecomposition2020}. Specifically, in our case it allows measuring three-waves coupling in the form of triad interactions present in the time-evolving schlieren dataset. 
The methodology relies on the assumption that the compressible Navier-Stokes equations can be considered as dynamical system driving the evolution of the state $\pmb q(\pmb x,t) = \pmb q_0(\pmb x) + \varepsilon \pmb q'(\pmb x, t)$, with a finite $\varepsilon$ value, such that expanding the equations into first to third order terms and retrieving the baseflow $\pmb q_0$ equation leads to, 
\begin{align}
    \frac{\partial \pmb q'}{\partial t} = \varepsilon\mathcal{J}\pmb q' + \varepsilon^2\mathcal{Q}(\pmb q', \pmb q') + \varepsilon^3\mathcal{T}(\pmb q', \pmb q', \pmb q'). \label{eq:nl-ns}
\end{align}
Where $\mathcal{J}$ is the linearised Navier-Stokes operator, termed the Jacobian. The operators $\mathcal Q$ and $\mathcal T$ respectively represent the quadratic and third-order non-linearities in the compressible NS equations. In this study, we investigate the flow in the initial stages of the transition regime where small disturbances $\pmb q'$ evolve around a steady fixed point $\pmb q_0$ such that $\pmb q = \pmb q_0 + \epsilon \pmb q'$. Considering $\varepsilon$ to be small but finite, leads to the growth of non-linearities in equation~\ref{eq:nl-ns}. However, recalling that we only look at the initial stages of transition, we consider the third-order non-linearities to be negligible $\mathcal T(\pmb q, \pmb q, \pmb q) \approx 0$, and we only retain the quadratic interactions $\mathcal{O}(\epsilon^2)$ described by $\mathcal{Q}$ under the form of triad interactions defined as,
\begin{align}
    f_i \pm f_j \pm f_k = 0. \label{eq:triad}
\end{align}
Therefore, the BMD seeks spatio-temporal coherent structures that follow a triad interaction as defined in equation~\ref{eq:triad}. Analogous to the bicoherence estimator \citep{liiCrossBispectrumComputationVariance1981,kimDigitalBispectralAnalysis1979a}, the BMD seeks for spatial coherent structures $\hat{\pmb q}(\pmb x, f)$ in the frequency space that maximise an expectation operator $S_{qqq}$,
\begin{align}
    S_{qqq}(f_i,f_j) = \lim_{T\rightarrow\infty}\frac{1}{T}E\left[\hat{\pmb q}^*(f_i)\hat{\pmb q}^*(f_j)\hat{\pmb q}(f_{i+j})\right]. \label{eq:bicoherence}
\end{align}
Which, for a multidimensional and discrete snapshot $\pmb{\hat{s}}$ at a frequency $f$ such as defined for the SPOD in Sec.\ref{sec:spod}, can be written in a discrete form,
\begin{align}
    b(f_i, f_j) = E\left[\hat{\pmb s}^H_{i\circ j}\mathsfbi W \hat{\pmb s}_{i+j}\right] \label{eq:vec-bic}
\end{align}
With $\mathsfbi W$, a weight matrix (identity in the case of schlieren images).
Using the bicoherence expression of equation~\ref{eq:vec-bic}, the BMD is based on two optimal linear expansions of Fourier modes over $N_{blk}$ realizations of the dynamics, with a common set of coefficients $\pmb a_{l}\in \mathbb C^{N_{blk}\times 1}$,
\begin{align}
    \pmb \phi^{[l]}_{i \circ j} = \hat{\mathsfbi Q}_{i \circ j}\pmb a_l &&
    \pmb \phi^{[l]}_{i + j} = \hat{\mathsfbi Q}_{i + j} \pmb a_l. \label{eq:bs-vecs}
\end{align}
Where $\pmb\phi^{[l]}_{i \circ j}(\pmb x, f_i, f_j)$ is the cross-frequency field representing the region of phase-alignment between frequency components of the triad. The vector $\pmb\phi^{[l]}_{i + j}(\pmb x, f_{i+j})$ is a bispectral mode that can be interpreted as the physical structure resulting from the interaction of frequencies $(f_i, f_j)$ \citep{schmidtBispectralModeDecomposition2020}. We seek for the unit norm expansion $\Vert\pmb a_1\Vert=1$, that maximises the Rayleigh quotient of the bispectral operator $\mathsfbi B$, which in turn defines its spectral radius and therefore gives its largest eigenvalue $\lambda_1(f_i, f_j) \in \mathbb{C}$.
\begin{align}
    \lambda_1 =  \left|\frac{\pmb a^H \mathsfbi B \pmb a}{\pmb a^H \pmb a}\right|, 
    \quad \text{with} \quad
    \mathsfbi B(\pmb x, \pmb x', f_i, f_j) = \frac{1}{N_{blk}} \hat{\mathsfbi Q}_{i \circ j}^H \mathsfbi W \hat{\mathsfbi Q}_{i + j}.
\end{align}
The complex bispectrum $\lambda_1(f_i, f_j)$ represents the amount of bispectral correlation at a given frequency doublet for additive or subtractive interactions depending on the region considered in the $(f_i, f_j)$ space.
Furthermore, using the vectors $\pmb \phi_{i \circ j}$ and $\pmb \phi_{i + j}$ defined in equation~\ref{eq:bs-vecs}, the spatial support of an active region of non-linear interaction can be observed by writing the interaction map,
\begin{align}
    \pmb \psi_{i,j}(\pmb x, f_i, f_j) = \left|\pmb \phi_{i \circ j} \circ \pmb \phi_{i + j} \right|. \label{eq:bmd-summary}
\end{align}
Using the complex bispectrum $\lambda_1$, the bispectral modes $\pmb \phi_{i + j}$ and the interaction map $\pmb \psi_{i,j}$, the triad interactions data can be extracted from the schlieren dataset.
Again, the algorithm is implemented using a distributed memory parallelisation to speed up the computation and spread the data on a large cluster.

\section{Flow Conditions Overview} \label{sec:expe-overview}



\begin{figure}
    \centering
    \begin{subfigure}{0.47\textwidth}
        \centering
        \includegraphics[width=\textwidth]{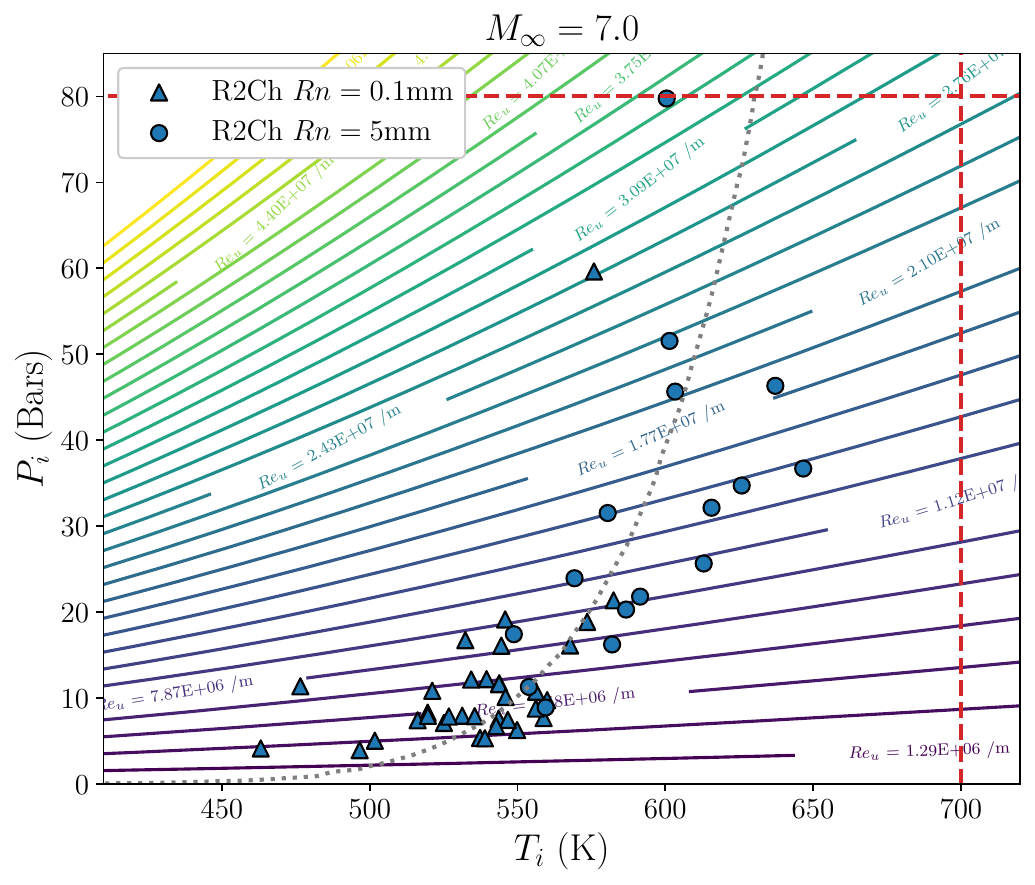}
    \end{subfigure}
    \hfill
    \begin{subfigure}{0.45\textwidth}
        \centering
        \setlength{\tabcolsep}{6pt} 
        \begin{tabular}{c c c c}
            $T_i$ [K] & $P_i$ [Bar] & $\Rey_\infty \times 10^{-6}$ & $M_\infty$ \\ \hline
            538.98 & 5.31 & 2.77 & 7 \\
            558.84 & 7.68 & 3.77 & 7\\
            550.28 & 7.59 & 3.82 & 7\\
            531.30 & 7.94 & 4.24 & 7 \\
            519.68 & 7.93 & 4.40 & 7 \\
            560.04 & 9.75 & 4.77 & 7 \\
            543.79 & 11.61 & 5.96 & 7 \\
            566.12 & 12.62 & 6.06 & 7 \\
            476.54 & 11.36 & 7.31 & 7 \\
            532.35 & 16.73 & 8.91 & 7 \\ \hline
        \end{tabular}
    \end{subfigure}
    \caption{Summary of all the $M_\infty=7.0$ flow conditions considered in the experimental campaign and the corresponding flow conditions used in this paper.}
    \label{fig:expe_matrix_and_conditions}
\end{figure}

To explore the transitional dynamics of the flow around the CCF12 geometry, runs were performed at a Mach number of 7 across a large range of Reynolds numbers. This Reynolds number $\Rey$ is defined for a length of 1 meter.
Figure \ref{fig:expe_matrix_and_conditions} summarises these runs by plotting each of them in the stagnation pressure and stagnation temperature space $(T_i, P_i)$ over a map of the associated Reynolds numbers. Runs for sharp nosetip are depicted with triangle markers. This data-set spans most of the Reynolds number and stagnation temperature range of R2Ch at both Mach numbers. A clear increasing trend in the runs stagnation temperature proportional to the stagnation pressure is noted and is inherent to the way  the R2Ch wind-tunnel was operated for this particular campaign which is related to the thermal inertia of the heated and compressed air circuit.
Detailed information about all the runs presented in this article is presented in table \ref{fig:expe_matrix_and_conditions}.


\subsection{Flow topology evolution with Reynolds number}

\begin{figure}
    \centering
    \includegraphics[width=\textwidth]{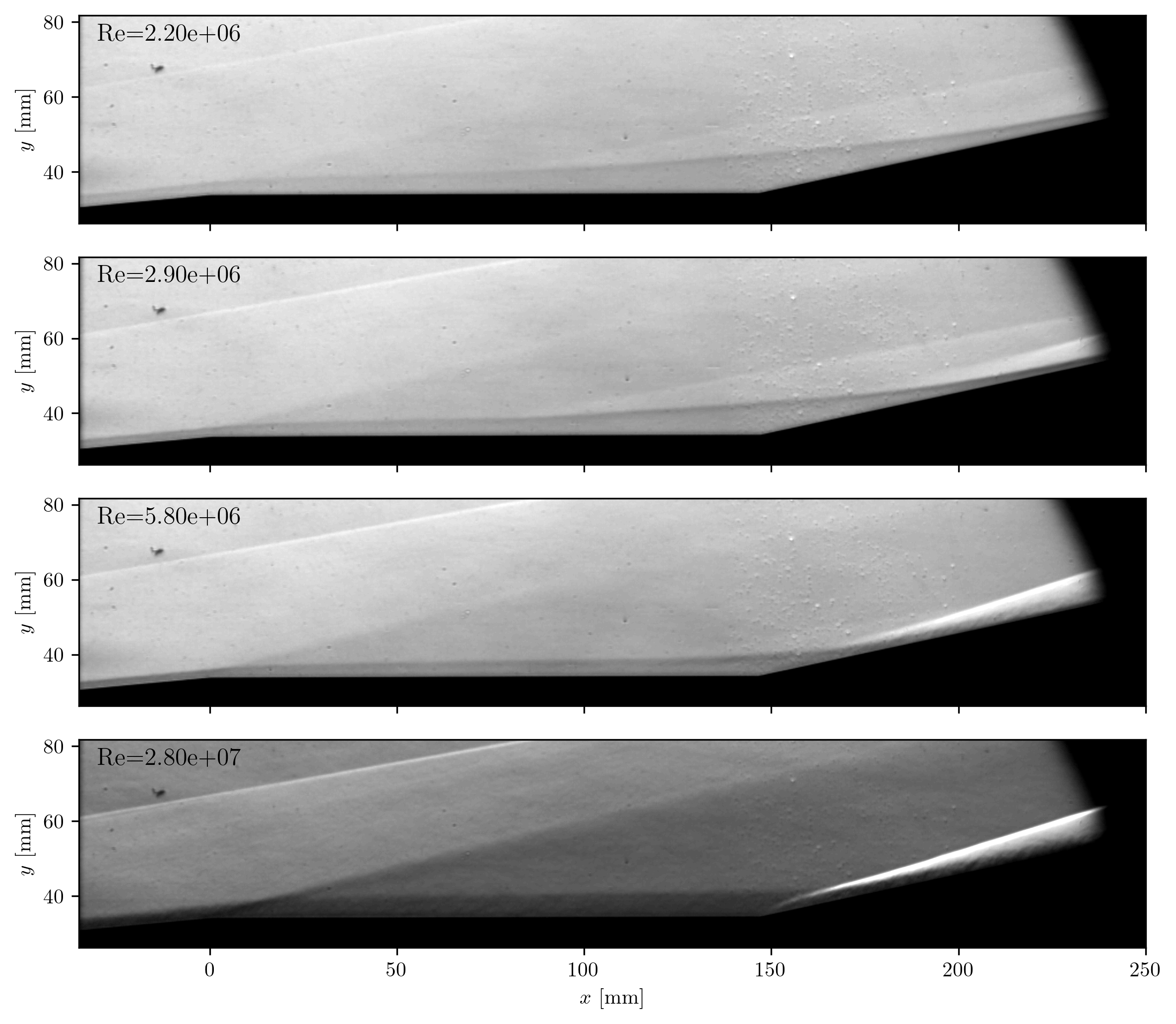}
    \caption{Mean strioscopy of the flow around CCF12 at increasing Reynolds number}
    \label{fig:mean_strio}
\end{figure}
\begin{figure}
    \centering
    \includegraphics[width=\textwidth]{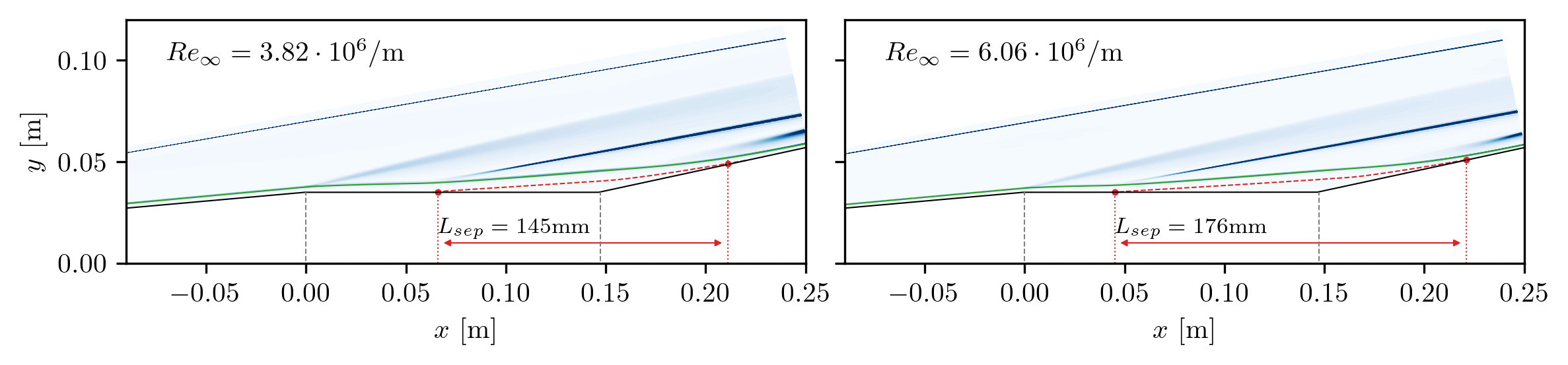}
    \caption{Numerical baseflows schlieren for selected conditions. Boundary-layer height (\ccontline{mplgreen}), recirculation bubble with separation and reattachment points (\cdashedline{mplred}) }
    \label{fig:numerical-bf}
\end{figure}
\begin{figure}
    \centering
    \includegraphics[width=0.8\textwidth]{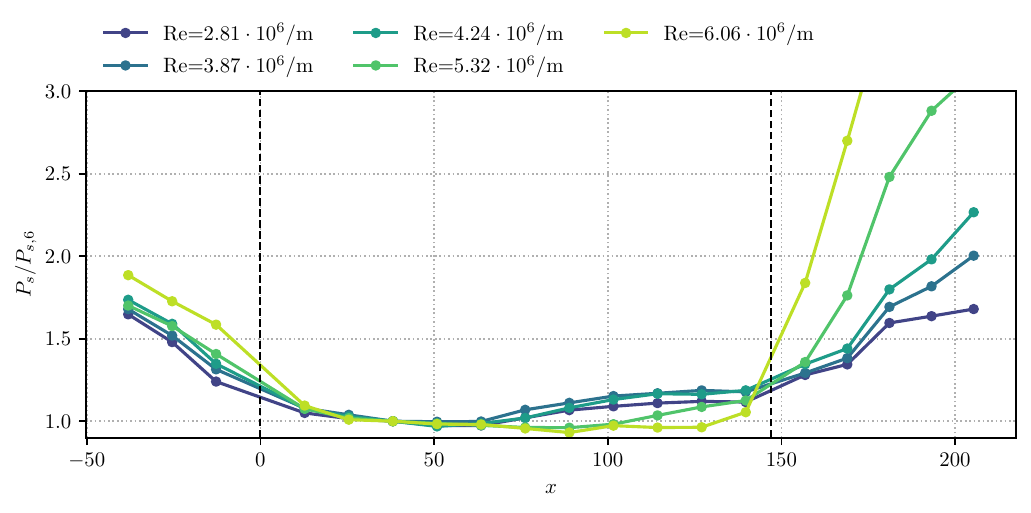}
    \caption{Mean pressure distribution along the geometry showing the dependence of the separation length to the Reynolds number. Dashed lines separate the cone, cylinder and flare regions}
    \label{fig:static-trends}
\end{figure}

\begin{table}
    \centering
    \begin{tabular}{ccccccccccc}
        \hline \hline   
        Bubble & $M_\infty$ & $Re_\infty \times10^6$ (m\textsuperscript{-1}) & $u_\infty$ (m/s) & $\rho_\infty$ (kg/m$^3$) & $T_\infty$ ($K$) & $T_0$ ($K$) & $T_\text{wall}$ ($K$) \\ \hline
        Large & 7.0 & 3.82 & 1001.69 & 0.01253 & 50.9537 & 550.3 & 300.0  \\
        Small & 7.0 & 6.06 & 1015.97 & 0.02026 & 52.4167 & 566.1 & 300.0  \\
        \hline \hline
    \end{tabular}
    \caption{\label{tab:flow-conditions} Detailed flow conditions used for the computations}
\end{table}

Previous studies have shown that, for transitional SBLI, there is a strong coupling between the separated laminar region and the transition process. This coupling is due to the fact that the transition process is strongly influenced by the incoming disturbance level in the boundary layer and the farfield \citep{marxenEffectSmallamplitudeConvective2011, lugrinTransitionScenarioHypersonic2021,benitez2025separation}.
This phenomenon is observed here at $M_\infty=7$ as the Reynolds number is increased. The figure~\ref{fig:mean_strio} highlights this effect by showing mean schlieren images of the separation bubble for different Reynolds numbers. A clear reduction of the size of the separated region can be seen as the Reynolds number increases and the separated region undergoes transition to turbulence. 
These experimental measurements contrast with the prediction of the laminar baseflow from a numerical computation using the framework described in section \ref{sec:num-tools} and visible in figure~\ref{fig:numerical-bf}. In the computation, the separated region (highlighted with the red arrows) is found to be continuously growing between Reynolds numbers of $\Rey=3.82\times 10^6$ and $\Rey=6.06\times 10^6$. Exact flow conditions of the simulated run are presented in Tab.~\ref{tab:flow-conditions}.
The recirculation bubble shrinking is further quantified by looking at the static pressure profile at the wall as the Reynolds number increases. The pressure trends are presented in figure \ref{fig:static-trends}, where the pressure profiles coloured by the Reynolds number show a clear reduction in length of the typical pressure plateau of the separated region. The separation point is located near the point where the non-dimensional pressure departs from 1 and starts to increase towards the plateau. The reattachment point is located where the pressure starts to increase from the plateau value.
The reduction occurs with both a downstream displacement of the separation point and an upstream displacement of the reattachment point. In the considered runs, the separation length initially increases between $\Rey=2.81\times 10^6$ and $\Rey=3.87 \times 10^6$ and then gradually decreases between $\Rey=4.24\times 10^6$ and $\Rey=6.06\times 10^6$. After this Reynolds number, the separation is not perceptible anymore in the static pressure measurements. The flow dynamics at $\Rey=3.87 \times 10^6$ and $\Rey=6.06 \times 10^6$ therefore appear to be the two runs of interest for the study of the separated region as they correspond to conditions where the bubble starts shrinking or nearly disappears, respectively.
This reduction is believed to be linked to the stronger amplification of the instabilities \citep{marxenEffectSmallamplitudeConvective2011, lugrinTransitionScenarioHypersonic2021}, and thus a faster transition caused by the increase in Reynolds number. The nature of these instabilities found around the CCF12 geometry is discussed in the next section.




\subsection{Global Resolvent analysis of the flow}
%
\begin{figure}
    \centering
    \begin{subfigure}[b]{0.495\textwidth}
         \centering
         \includegraphics[width=\textwidth]{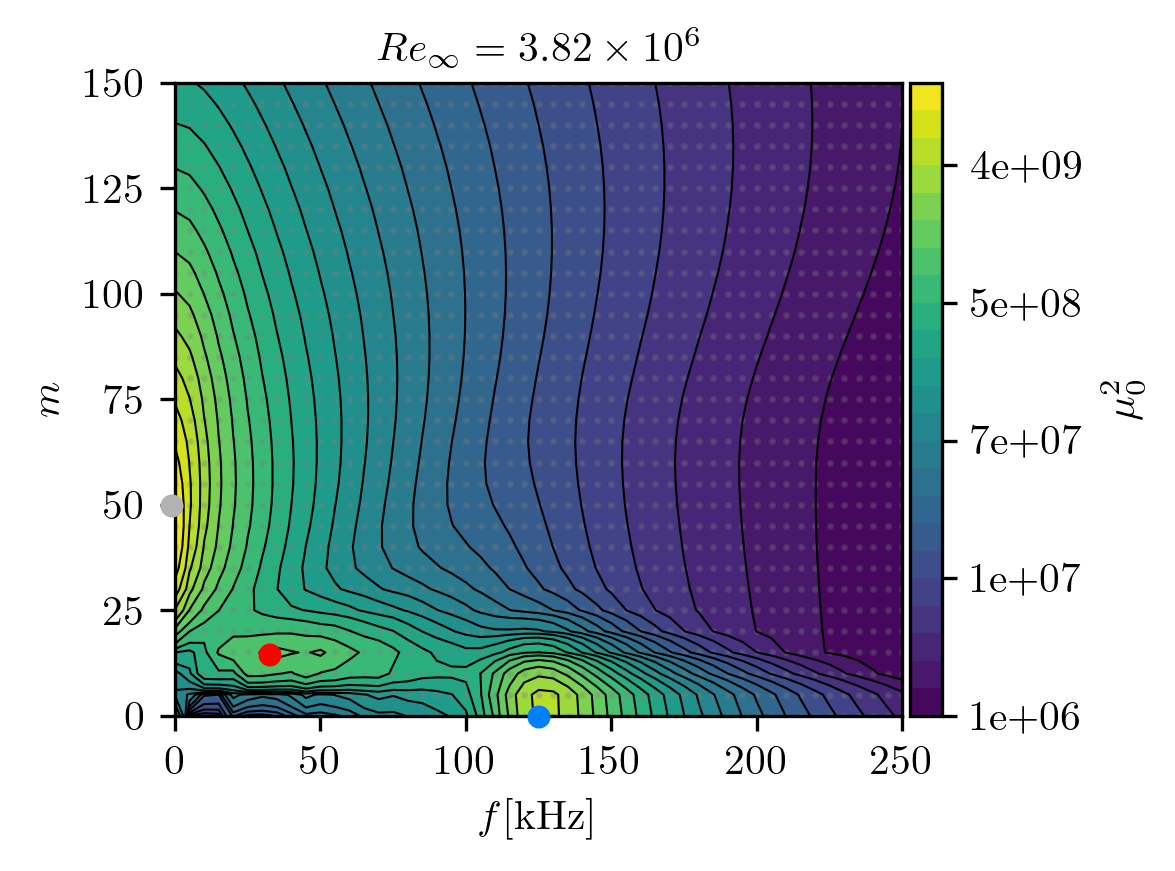}
         \caption{Optimal gain map for $\Rey_\infty = 3.82 \times 10^6$}
         \label{subfig:gain038}
     \end{subfigure}
     \hfill
     \begin{subfigure}[b]{0.495\textwidth}
         \centering
         \includegraphics[width=\textwidth]{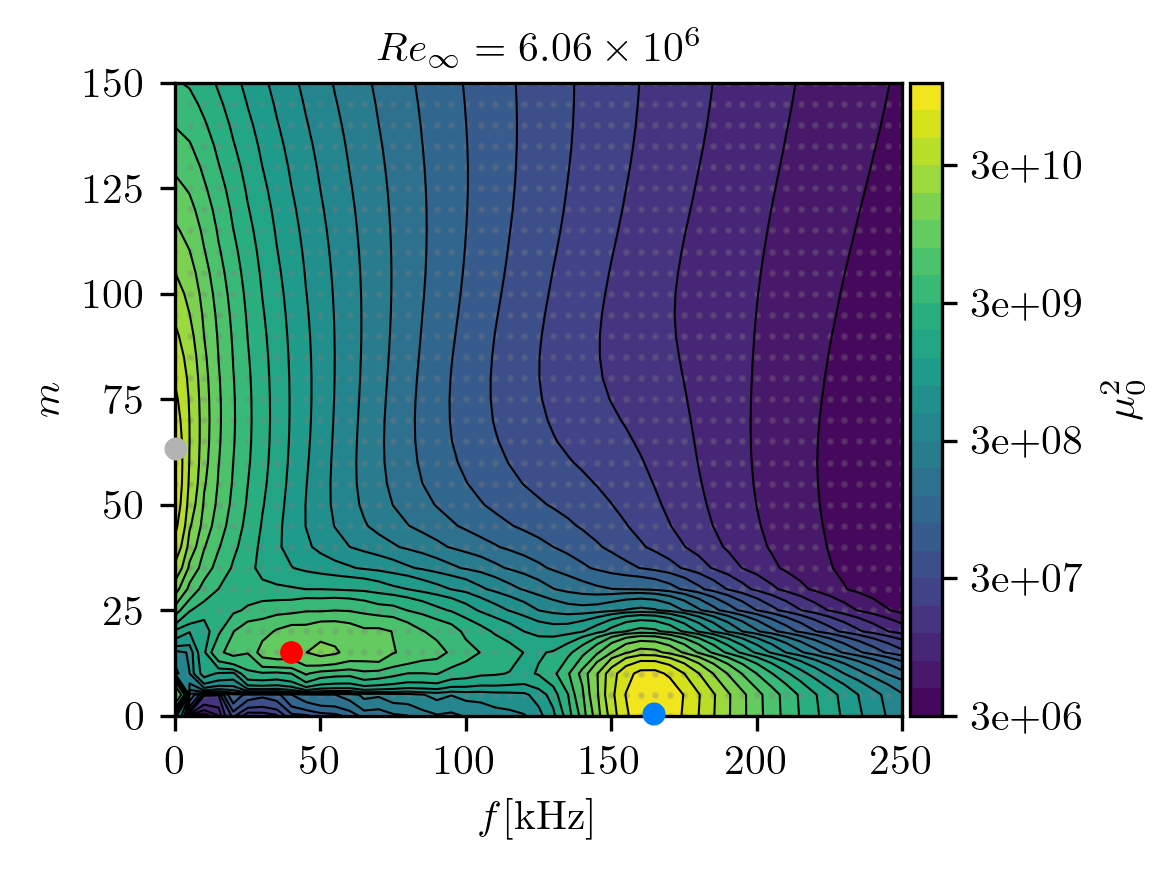}
         \caption{Optimal gain map for $\Rey_\infty = 6.06 \times 10^6$}
         \label{subfig:gain060}
     \end{subfigure}
     \\
    \begin{subfigure}[b]{\textwidth}
        \centering
        \begin{tikzpicture}
            \node[anchor=south west, inner sep=0] (image) at (0,0) {\includegraphics[width=\textwidth]{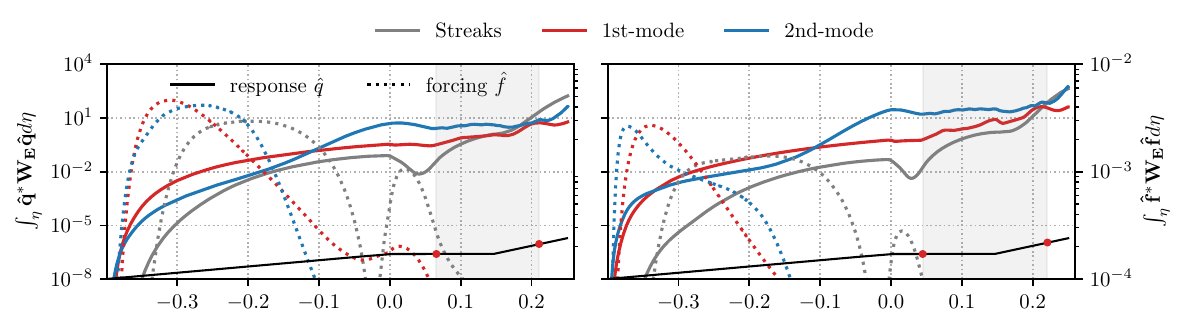}};
            \begin{scope}[x={(image.south east)}, y={(image.north west)}]
                \node[anchor=south, font=\small] at (0.3, -0.075) {$x$ [m]};
                \node[anchor=south, font=\small] at (0.73, -0.075) {$x$ [m]};
            \end{scope}
        \end{tikzpicture}
        \caption{Optimal responses and forcings energy for the highlighted amplification peaks in the gain maps with coloured markers}
        \label{subfig:chu_nrj_modes}
    \end{subfigure}
    \caption{Overview of the optimal gain and optimal modes energy evolution as the Reynolds number increases. Small grey dots on the gain maps indicate the discretisation of the frequency-azimuthal wavenumber space}
    \label{fig:resolvent_gains}
\end{figure}
An overview of the growing non-modal instabilities in this flow is provided in figure~\ref{fig:resolvent_gains}. Optimal gain maps from the Resolvent analysis around the full object display the leading amplification mechanisms.
The main hypersonic instabilities, such as the first- and second-mode are directly retrieved by the optimal gain maps in figures~\ref{subfig:gain038} and \ref{subfig:gain060}. For the selected Reynolds numbers of $\Rey_\infty=3.82 \times 10^6$ and $\Rey_\infty=6.06 \times 10^6$, the first-mode is visible as moderately oblique waves having a peak of gain between $f_{1st}=20$ and $f_{1st}=110$kHz with azimuthal wavenumbers $m\in[10,25]$.  The second mode is visible as a strong gain peak for mostly planar waves $m\in[0,10]$ and frequencies $f_{2nd} \in [100,250]$kHz. Finally, a strong amplification of steady to low frequencies and nearly streamwise waves is visible on the left side of the gain maps for $m\in[25,100]$, this gain peak corresponds to low frequency streamwise streaks having their highest amplification for $f=0$kHz. Considering these two gain maps it should also be noted that the most linearly amplified instability mechanism switches between the streamwise streaks and the second-mode as the Reynolds goes from  $\Rey_\infty=3.82 \times 10^6$ to $\Rey_\infty=6.06 \times 10^6$. 

The optimal Resolvent modes at the peaks highlighted with coloured dots in figures~\ref{subfig:gain038} and \ref{subfig:gain060} are discussed in figure~\ref{subfig:chu_nrj_modes}. In the two plots, optimal forcings and response energy distributions along the geometry are shown. The forcings distributions highlight regions where the receptivity process is likely to occur (following equation~\ref{eq:receptivity}). First- and second-mode waves show more sensitivity to incoming disturbance on the cone upstream region. This receptivity region is even more localised as the Reynolds number increases, additionally the first-mode receptivity region is found to get larger than the second-mode on the cone. On the other hand, the streaks receptivity remains fairly unchanged as the Reynolds number increases, with only a slight decrease of the optimal forcing amplitude close to the separation, highlighted in grey. The associated optimal responses energy distribution helps identifying the leading response mechanisms in each region of the flow. For both Reynolds number, the second-mode is the dominating mechanism at the end of the cone, followed by the first-mode instability and the streaks, each of these modes are separated by at least one order of magnitude in amplitude. After the expansion at $x=0.0$ both the second-mode and the streaks show a damping trend which is more pronounced for the streaks and increases with the Reynolds number. The first-mode remains amplified and keeps amplifying in the separated region, with growth peaks close to the curvature regions of the separation and reattachment, suggesting an amplifying mechanisms linked to curvature at these stations. Similarly, the streaks are strongly amplified at the separation and reattachment. For this latter mode, the growth mechanisms can be linked to a mix of lift-up and Görtler mechanisms in the separation region \citep{Song_Hao_2025}. Consistent with previous studies \citep{paredes2022boundary,caillaud2025separation}, the second-mode remains nearly neutral in the separation region and is strongly amplified after reattachment. However, while it remains neutral, it should be observed that this mode still dominates the two other instabilities in energy up to the beginning of the flare, suggesting its important role in the transitional dynamics of the separated region. Finally, at the end of the geometry, the dominating modes are found to be the streaks for the lower Reynolds number and both streaks and second-mode for the higher Reynolds number, where they display similar amplitudes.
Further details about those instabilities around the CCF12 geometry can be found in \citet{caillaud2025separation}.

\subsection{Reynolds trends in unsteady measurements}

\begin{figure}
    \centering
    \includegraphics[width=\textwidth]{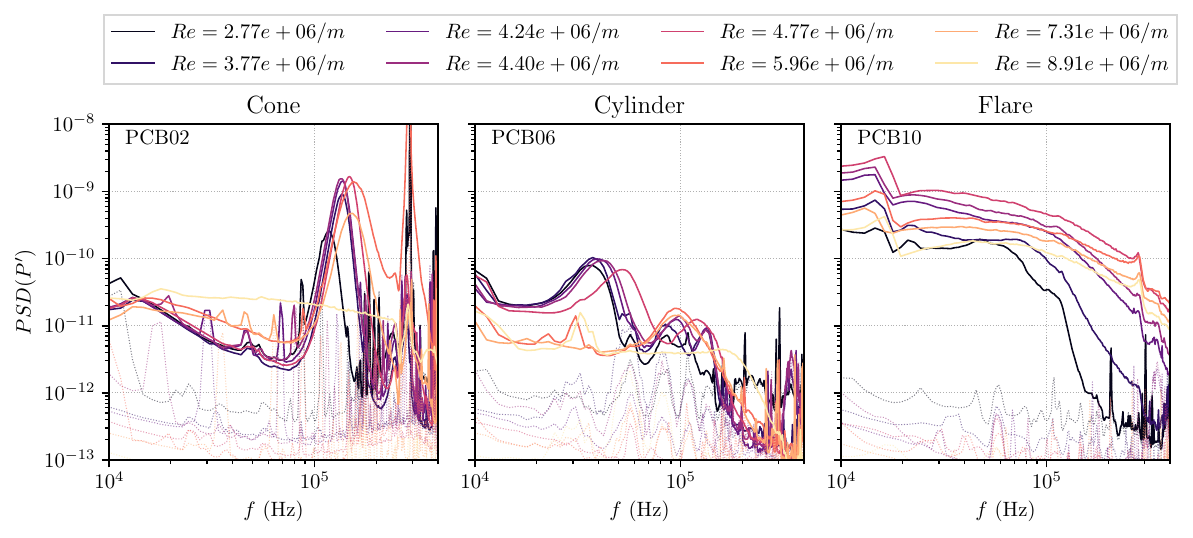}
    \caption{Pressure power spectral densities from PCB measurements on the cone, cylinder and flare for different Reynolds number. The sensors noise floor is shown in dotted lines.}
    \label{fig:PSD-re}
\end{figure}

In relation with the global stability analysis, figure \ref{fig:PSD-re} presents part of the experimental data in the form of wall pressure power spectral densities for the three main regions of the object: cone, cylinder and flare. In this figure, the main wall-pressure amplification peaks related to the three regions of the geometry can be observed.
First, the second mode is highlighted in figure \ref{fig:PSD-re}a with well-marked peaks visible between $115$kHz and $190$kHz.
On the cylinder clear peaks in the range of $25$kHz to $40$kHz also appear and have been linked with first- or shear-layer modes \cite{benitezInstabilityMeasurementsAxisymmetric2020}.
On the flare section, the spectra are rapidly showing turbulent signature with a typical broadband energy content. 

Focusing on the second-mode instability, the amplification of this instability is already visible at the lowest Reynolds number of $\Rey = 2.77 \times 10^6$ and seems to saturate around $\Rey=6.0\times 10^6$ at the end of the cone.
The second-mode signature remains visible on the cylinder with lower amplitude peaks around $150$kHz between $\Rey=4\times 10^6$ and $\Rey=5.5\times 10^6$, this decrease in amplitude is caused by the expansion fan at the cone-cylinder junction. 
At the end of the flare, as expected for turbulent flow, the second-mode signature is not clearly visible for all the considered Reynolds numbers.

Contrary to the second mode, the first-mode instability does not display any pressure signature at the end of the cone.
This behaviour is consistent with the nature of the instability and the previous experimental results \citep{butlerTransitionalHypersonicFlow2022,benitezInstabilityTransitionOnset2023}. However, clearly visible peaks in the spectra on the cylinder for $\Rey_\infty \in [2.77 \times 10^6, 4.77 \times 10^6]$ may be related to a lower frequency mechanisms, such as the first-mode. This analysis will be the subject of a later section.
The lower-frequency wall-pressure signature shows a broad peak being one order of magnitude higher than the higher frequency second-mode at equivalent Reynolds number.
A moderate shift of frequency is noticed across the different conditions but remains restrained as the Reynolds number variation is relatively moderate between these runs. 
On the flare the low-frequency peak can be noticed for $\Rey_\infty = 2.77 \times 10^6$ with a spread hump however, this peak quickly saturates to a broadband signature at higher Reynolds number $\Rey>4\times10^6$.

It is noteworthy that the cylinder pressure spectra depict a more intricate pressure signature than the cone and the flare. 
In addition to the two first- and second-mode signatures, multiple pressure peaks are visible at intermediate frequencies, these peaks arise and vanish depending on the Reynolds number.
This suggests that these dynamics are related to the transitional process linked to specific bubble flow topologies, i.e. separation bubble lengths. This assumption is also addressed in more details in the following sections.

Finally, as depicted in figure~\ref{fig:PSD-re}, beyond a Reynolds number of $\Rey_\infty \geq 4.77 \times 10^6$, the lower frequency peak instability ceases to manifest on the cylinder, whereas the second mode experiences significant amplification on the cone. 
This goes in pair with the strong decrease in separated region length past this Reynolds documented figure \ref{fig:static-trends}.
Coupled with the observed additional spectral peaks, these trends in Reynolds numbers indicate a pathway to turbulence that transitions from a cylinder-dominated scenario for larger separations and lower Reynolds numbers to a cone-dominated one, with small or non-existent separation scenario for higher Reynolds numbers. This trend exhibits similarities with reentry trajectories where the transition front moves upstream along the vehicle's geometry as the altitude lowers and the Reynolds number increases.

Therefore, the next sections will set the focus on the two aforementioned scenarios. A first selected case at $\Rey=3.82\times 10^6$, marking the beginning of the bubble size reduction and transition on the flare. A second case at $\Rey=6.06\times 10^6$ where the separated region nearly vanishes and the wall pressure spectra immediately display a broadband signature on the flare. This latter case is investigated in what follows.


\section{Transition with a small separation from cone instabilities} \label{sec:small-sep}

The regime where transition is driven by the amplification and interactions of second-mode waves coming from the cone is discussed first as it offers a more direct interpretation of the transition stages. 
The case at $\Rey_\infty = 6.06\times 10^6$ highlights the main aspects of this second-mode dominated transition. 
As shown in figure \ref{fig:static-trends}, this case shows nearly no separation as only the last pressure tap before the cylinder shows a slight increase in pressure.
As such, this regime is mainly about attached boundary layer instabilities and one can expect the flow to be very close to being transitioned at the beginning of the SBLI.
The route to turbulence will be discussed by comparing measurements obtained with PCB wall-pressure sensors, time-resolved high-speed schlieren imagery and global stability analysis. 
It can be noted here that for that case all the PCB sensors are located on regions where the flow is attached.
%
%
\subsection{Second mode waves}
\begin{figure}
    \centering
    \includegraphics[width=\textwidth]{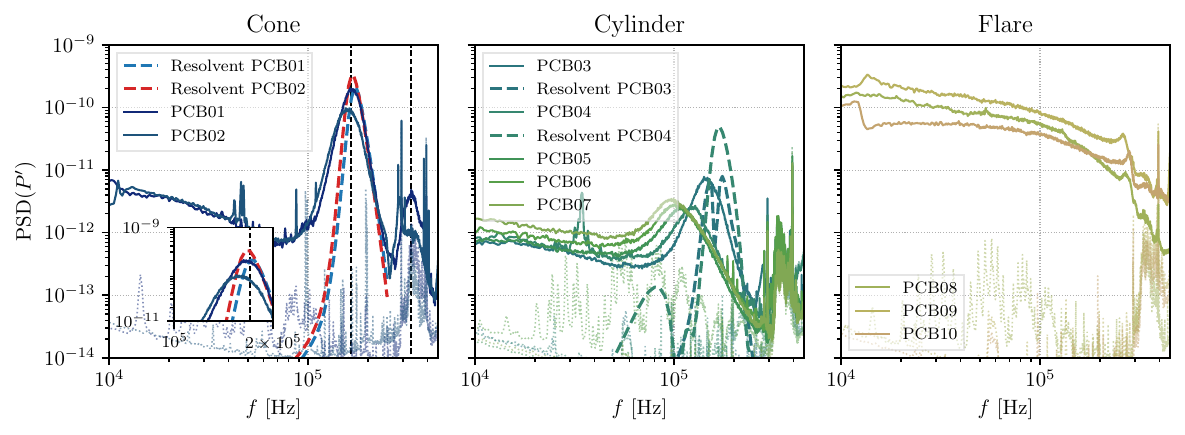}
    \caption{PCB wall pressure spectrum evolution along the geometry for $\Rey_\infty=6.06\times 10^6$}
    \label{fig:case5469-pcb-along-geom}
\end{figure}

An overview of the wall-pressure frequency spectra along the cone for these conditions is shown in figure~\ref{fig:case5469-pcb-along-geom} for sensors in the three main region of the model, the cone, the cylinder and the flare, respectively. All the spectra of the cone and cylinder exhibit at least one large peak in the range $f\in[90, 200]$~kHz, representative of the second Mack mode instability. 
However, on the flare, a broadband spectrum, typical of the turbulence regime, is visible for $f\leq 2 \times 10^5$~Hz and suggests that the flow has already undergone transition at the beginning of the flare.

The amplified peaks visible on the cone pressure fluctuation spectra in figure~\ref{fig:case5469-pcb-along-geom} are identified from the Resolvent analysis performed in figure~\ref{subfig:gain060} as being second mode waves. 
For this identification, the Resolvent optimal responses are used to reconstruct a wall-pressure spectrum for the most amplified wavenumber $m=0$. Due to their linear nature, the Resolvent optimal responses have relative amplitudes.
Therefore, the pressure response is scaled by matching the amplitude of the most amplified optimal response to the PCB01 measurement at the second-mode frequency. Using this scaling, it clearly appears that the Resolvent analysis accurately predicts the second-mode frequency for this Reynolds number for PCB01.
However, the subsequent amplification of the optimal responses on the next sensor (PCB02, PCB03 and PCB04) of figure~\ref{fig:case5469-pcb-along-geom} does not match the actual amplification trends of the experiments. Added to the visible damping and spreading of the second-mode peak at the PCB02, shown in the zoomed inset, it indicates that the flow has entered the non-linear saturation regime at the end of the cone.
%

The evolution of these saturating waves is further detailed with the Cylinder spectra of figure~\ref{fig:case5469-pcb-along-geom} where the wall pressure spectrum for the PCBs on the cone and cylinder are shown. Specifically, the evolution of the saturating second-mode is visible from the continued spreading and dampening of the peaks. The direct comparison before the separation, to the linear prediction of the leading Resolvent mode discussed above, shows a clear mismatch in frequency and amplitude with the experimentally measured waves. The second mode peaks keep broadening along the cylinder indicating ongoing non-linear saturation of these waves. Finally, the flare spectra of figure \ref{fig:case5469-pcb-along-geom} only displays broadband energy content typical of turbulence at every downstream sensors, indicating that the boundary layer is far in the non-linear regime in this region of the flow.

\subsection{Optical measurements}

\begin{figure}
    \begin{subfigure}[h]{0.55\textwidth}
        \centering
        \includegraphics[width=\textwidth]{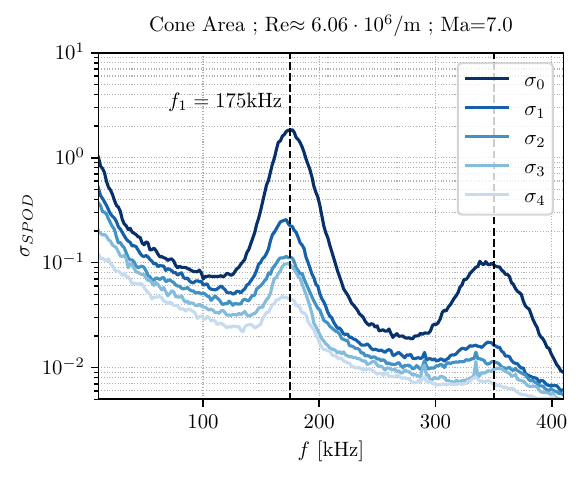}
        \caption{SPOD energy spectrum for $\sigma_0 > ... > \sigma_4$}
        \label{subfig:cone-spod-nrj}
        \centering
    \end{subfigure}
    \hfill
    \begin{subfigure}[h]{0.445\textwidth}
        \begin{subfigure}[b]{\textwidth}
            \centering
            \includegraphics[width=\textwidth]{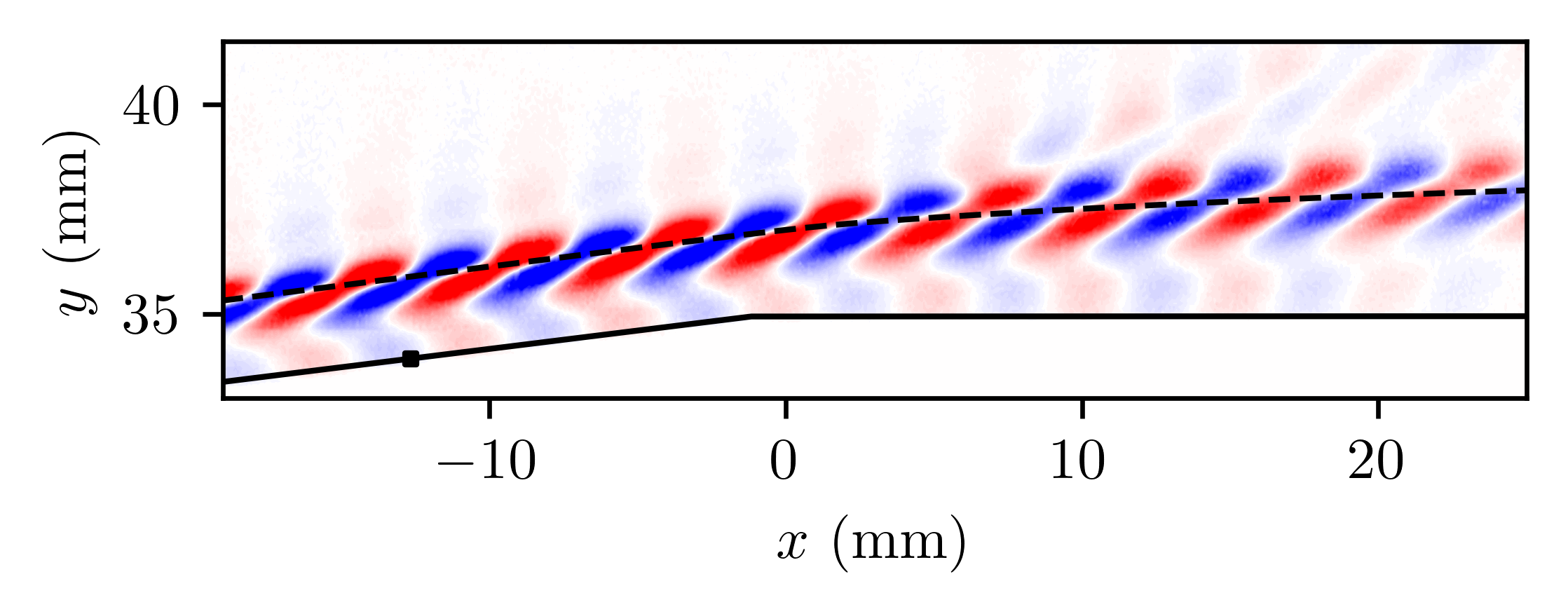}
            \caption{SPOD second-mode at $175$kHz}
            \label{subfig:spod-2nd}
        \end{subfigure}
        \begin{subfigure}[b]{\textwidth}
            \includegraphics[width=\textwidth]{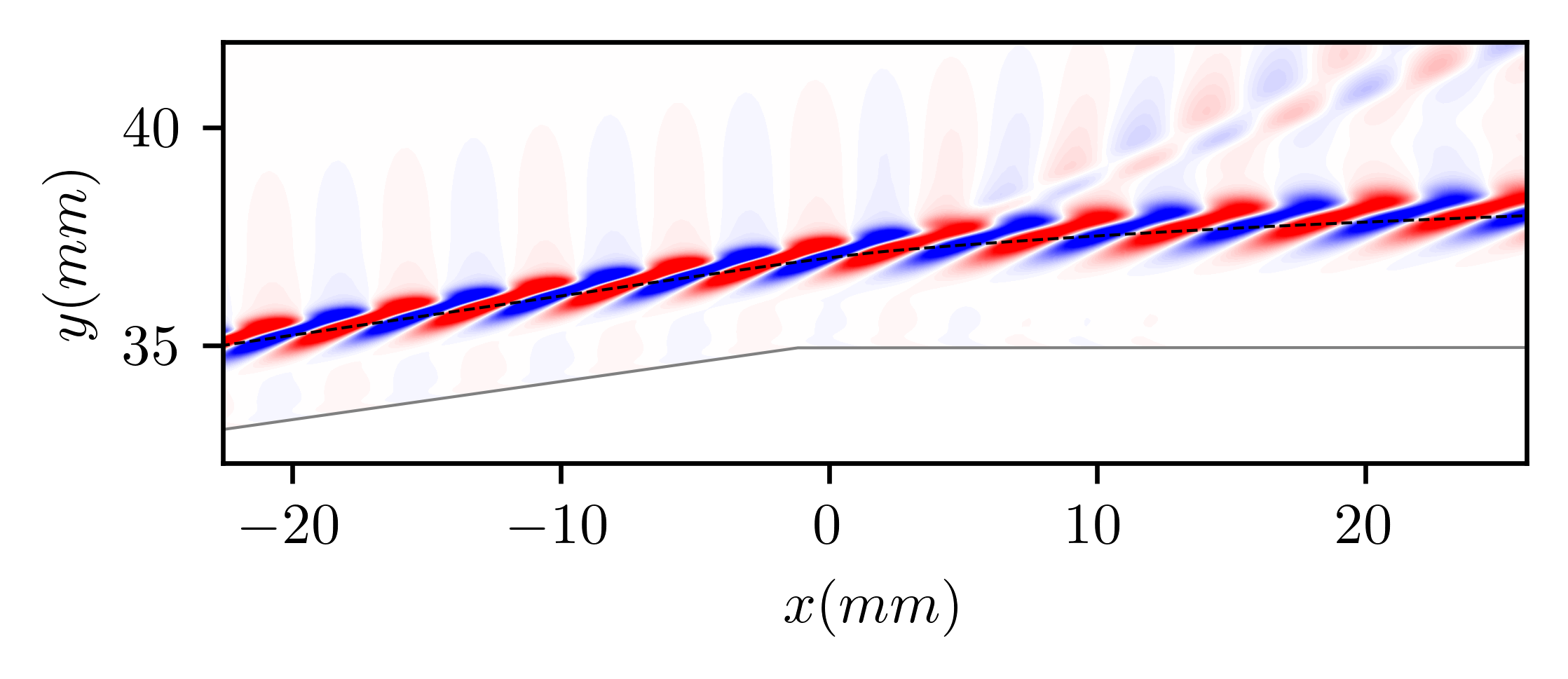}
            \caption{Resolvent second-mode at $f=175$kHz, $m=0$}
            \label{subfig:res-2nd}
        \end{subfigure}  
    \end{subfigure}

    \caption{SPOD results for the cone schlieren dataset at $\Rey_\infty=6.06\times 10^6$}
    \label{fig:spod-2nd-mode}
\end{figure}

The discussion on the pressure spectra peaks observed in figure~\ref{fig:case5469-pcb-along-geom} is completed by a SPOD and a BMD analysis of the high speed schlieren data gathered on the cone. A frame extending from the end of the cone to the beginning of the cylinder $-22 \leq x \leq 25$ is considered. 
The SPOD energy spectrum  is presented in figure~\ref{subfig:cone-spod-nrj}.
%

Two peaks with a substantial separation are clearly visible in the SPOD energy spectrum at $f_1=175$kHz and $f_2=350$kHz. The frequency $f_1$ is directly matching the peak observed on PCB02 for pressure fluctuation at the end of the cone in figure~\ref{fig:case5469-pcb-along-geom}, suggesting it to be linked with a second mode.
The second peak displays a lower energy level and seems to be matching a triad relation $f_1 + f_1 = 2 f_1$, leading to a super-harmonic. 
Such non-linear dynamics are discussed in the next section. 
For now, the discussion will focus on the comparison of the leading SPOD mode and the resolvent mode which are shown in figure~\ref{subfig:spod-2nd} and ~\ref{subfig:res-2nd}. 

For the leading peak at $f_1$, the SPOD mode clearly exhibits a coherent structure with a short streamwise periodicity, mainly localised along the line of maximum $\nabla \rho$ (black-dashed line) 
and close to the wall. 
This SPOD mode organisation is closely aligning with the Resolvent second-mode mode (figure~\ref{subfig:res-2nd}).
Few differences reside in the organisation of the most energetic regions, these differences can be attributed to three experimental effects: the distortion imposed by the parallel integration path of the schlieren light rays through the axisymmetrical flow, the superposition of different azimuthal modes and the onset of the non-linear regime. Nonetheless, both the Resolvent optimal response and SPOD modes exhibit similar shapes, with a visible wall-signature and acoustic energy being radiated outward by the expansion wave region.


\subsection{Discussion on the first non-linear stages}
\begin{figure}
    \centering
    \includegraphics[width=0.9\textwidth]{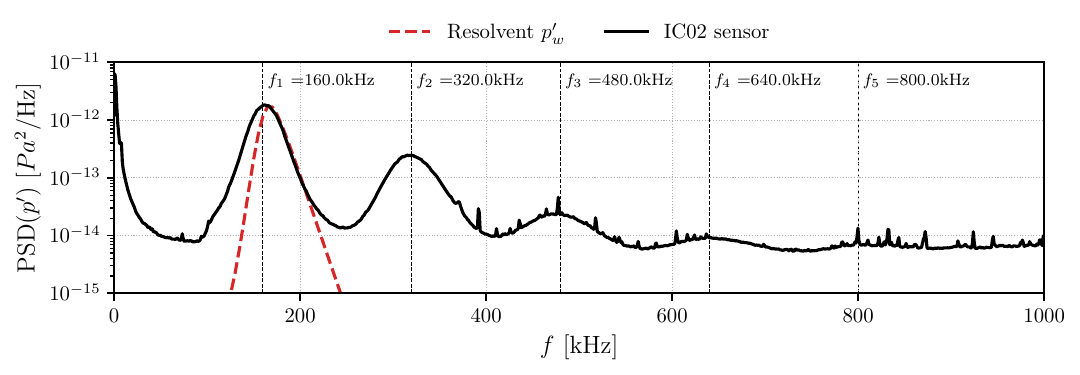}
    \caption{Pressure power spectral density measured with IC02 for $\Rey_\infty=6.06\times 10^6$ showing the multiple harmonics of second mode along with the linear wall-pressure prediction of the Resolvent operator at this station.}
    \label{fig:IC2_spectrum}
\end{figure}

The SPOD energy spectrum and the PCB measurements revealed a possible non-linear interaction of the second mode with the presence of two  peaks in the spectrum at the end of the cone region. 
To get a better understanding of this non-linear process, measurements at the same Reynolds number have been conducted with an IC2 sensor at the position of PCB02.
The resulting spectrum is presented in figure \ref{fig:IC2_spectrum} and displays multiple peaks from $160$kHz to $800$kHz in addition to the one already discussed in the PCB spectra. All those peaks are found at frequencies that are multiples of the fundamental frequency, which strongly suggests that they come from a self-interaction of second mode waves and the subsequent non-linearly produced harmonics. This assumption is further supported by the absence of additional peaks in the linear response of the Resolvent shown with a red dashed line. 

\begin{figure}
    \centering
    \begin{subfigure}[b]{0.49\textwidth}
         \centering
         \includegraphics[width=\textwidth]{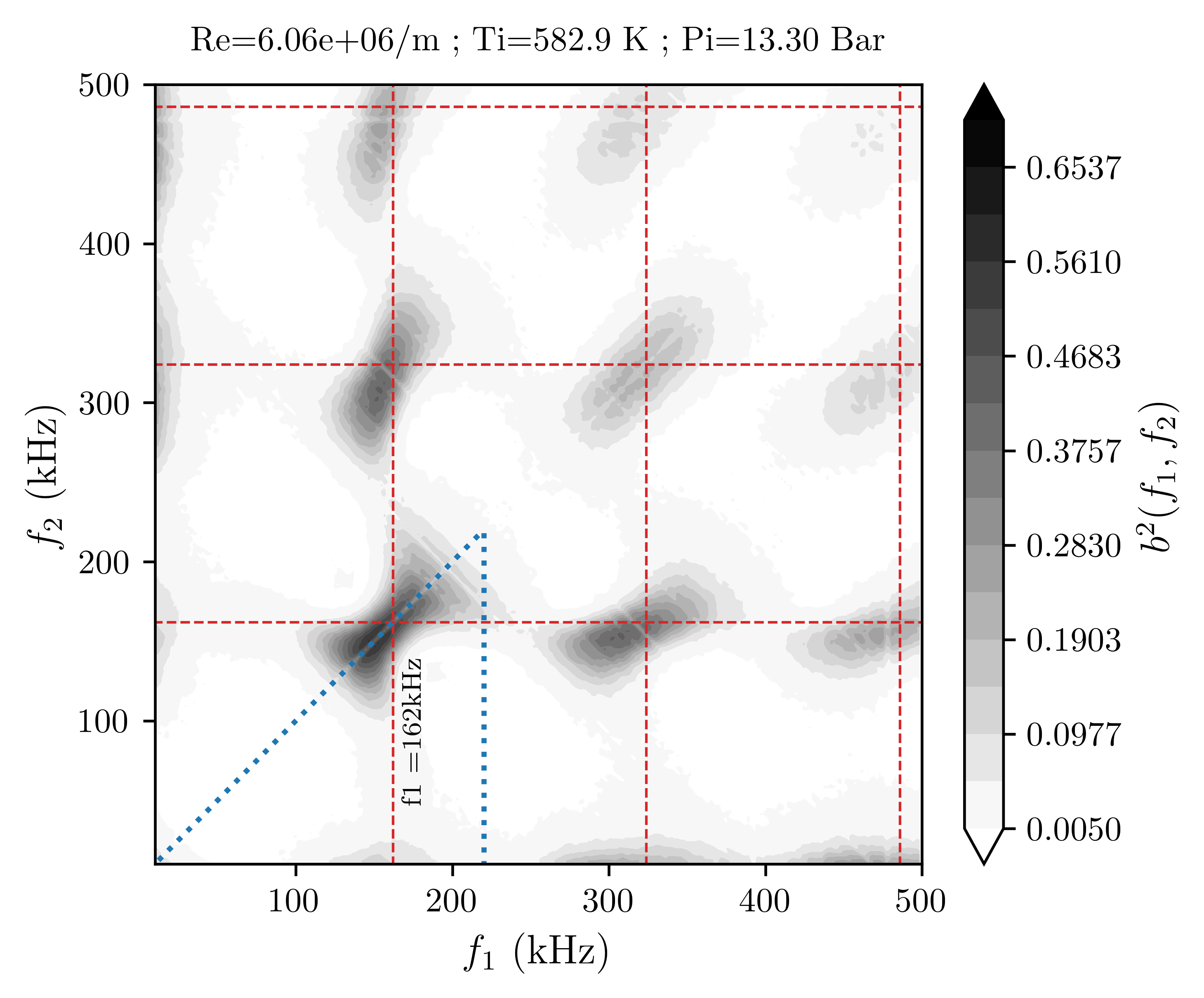}
        \caption{Bicoherence $S_{qqq}$ of IC202}
        \label{subfig:cone-bs}
    \end{subfigure}
    \hfill
    \begin{subfigure}[b]{0.49\textwidth}
        \centering
         \includegraphics[width=\textwidth]{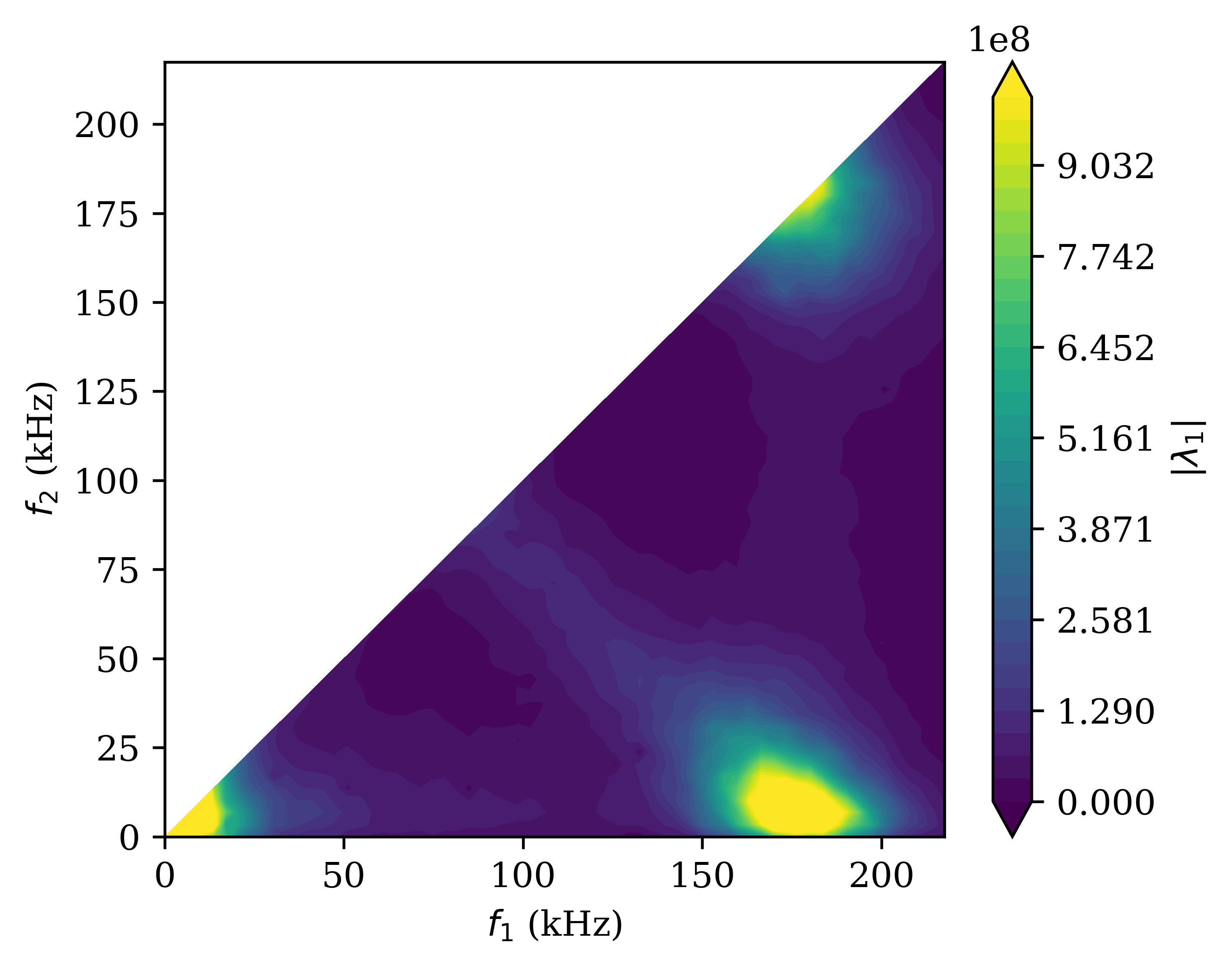}
         \caption{Complex bispectrum $\lambda_1$ from the schlieren}
         \label{subfig:cone-bmd}
     \end{subfigure}
    \caption{Bicoherence estimators from different metrologies in the cone-cylinder region. The BMD region is outlined as a blue triangle in the IC02 bicoherence map.}
    \label{fig:case5469_bicoherence_analysis}
\end{figure}

Those possible three waves interactions are further investigated by using the bicoherence estimator of equation~\ref{eq:bicoherence} for the measurements of IC202 and the complex bispectrum of the BMD, defined in equation~\ref{eq:bmd-summary}, for the high-speed schlieren dataset with the same field of view as the SPOD. 
Results from these two estimators are shown in figures ~\ref{subfig:cone-bs} and \ref{subfig:cone-bmd}.

In both figures, peaks of phase-correlations between signals at frequencies respecting a triad-interaction are visible. 
Specifically, in the IC202 bicoherence map which offers the greatest frequency range, a rich set of triad-interactions is clearly visible for increased multiples of the second-mode peak frequencies. 
Considering $f_1$, the second-mode frequency, the following possible three-waves interactions are visible in the bicoherence map of figure~\ref{subfig:cone-bs}. 
\begin{align}
    f_1+f_1&=2 f_1, & f_1 + 2f_1 &= 3 f_1, & f_1 + 3 f_1 &= 4 f_1, & 2 f_1 + 2 f_1 &= 4 f_1, \\
    f_1-f_1&=0,     & f_1 - 2f_1 &= f_1,   & f_1 - 3 f_1 &= 2 f_1, & 2 f_1 - 2 f_1 &= 0,  \\
    3 f_1 + 3 f_1 &= 6 f_1, & 3 f_1 + 2 f_1 &= 5 f_1, & 3 f_1 + f_1 &= 4 f_1, & 3 f_1 + 0 &= 3 f_1.
\end{align}
As the Bicoherence estimator of equation~\ref{eq:bicoherence} does not allow to obtain a causal directivity in the triad-interactions, both subtractive and additive three-waves relations have to be considered. The prevalence of one over the other is obtained through a context on the flow dynamics  \citep{bountinEvolutionNonlinearProcesses2008,craigNonlinearBehaviourMack2019}. Namely, the presence of a second and third super-harmonic on the IC202 frequency spectrum confirms the presence of additive triad-interactions $f_1+f_1=2 f_1$ and $f_1 + 2f_1 = 3 f_1$.
These interactions are further confirmed by considering the BMD complex bispectrum $\lambda_1(f_i, f_j)$ in figure~\ref{subfig:cone-bmd}. As discussed in Sec.~\ref{sec:bmd}, the complex bispectrum provides a directional information on the possible phase correlation (e.g. it distinguishes additive or subtracting interactions regions) \citep{schmidtBispectralModeDecomposition2020}. Looking at its value in the positive quadrant $(0\leq f_2 \leq f_1)$ allows observing additive energy transfers. Two leading interactions are visible at $f_1=175$~kHz,
\begin{align}
    0 + f_1 = f_1 && f_1 + f_1 = 2 f_1.
\end{align}
The former interaction indicates that the lower frequency content $0 \leq f \leq 20 kHz$ is transferring energy to the second mode. Such interactions usually corresponds to an extraction of the energy of the baseflow from the instability. The second interaction corresponds to a first super-harmonic generation.
\begin{figure}
    \centering
    \begin{subfigure}[b]{0.49\textwidth}
         \centering
         \includegraphics[width=\textwidth]{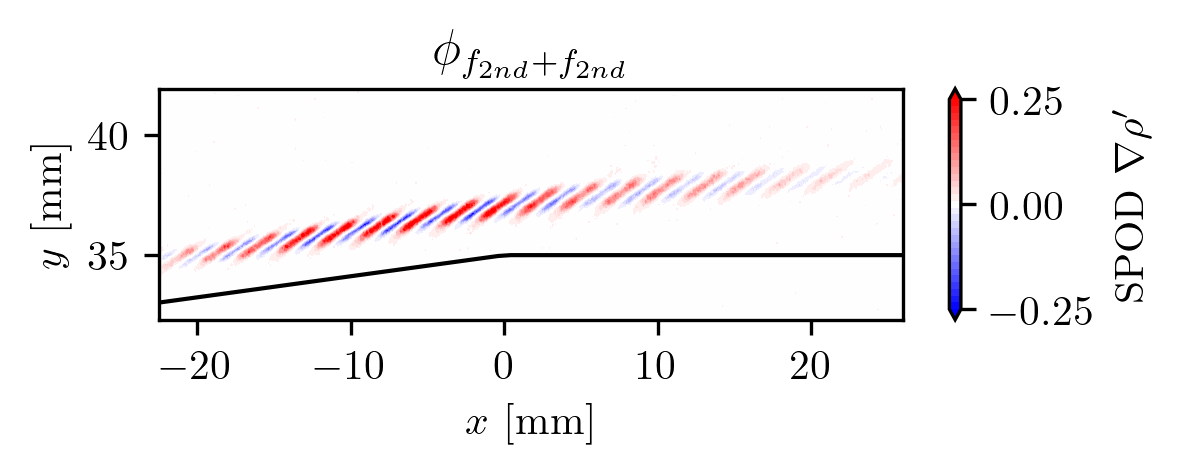}
        \caption{First-harmonic of the second mode}
        \label{subfig:harmonic}
    \end{subfigure}
    \hfill
    \begin{subfigure}[b]{0.49\textwidth}
        \centering
         \includegraphics[width=\textwidth]{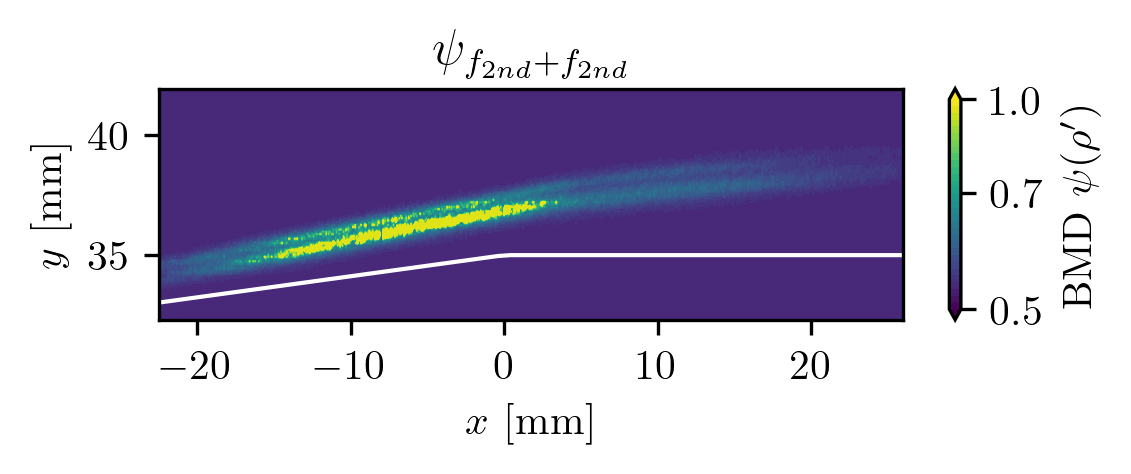}
         \caption{Interaction map}
         \label{subfig:interaction_map}
     \end{subfigure}
    \caption{First-harmonic of the second mode  $\pmb \phi_{i + j}$ and normalised interaction map $ \pmb \psi_{i,j}(x, y, f_i, f_j)$ of the triad interaction $f_{2nd}+f_{2nd}=2f_{2nd}$ from the BMD}
    \label{fig:harmonic_generation}
\end{figure}

More details on the generation of this latter second-mode harmonic are provided in figure~\ref{fig:harmonic_generation} where the BMD is used to retrieve the first-harmonic mode $\pmb \phi_{i + j}$ (figure~\ref{subfig:harmonic}) and the interaction map $ \pmb \psi_{i,j}(x, y, f_i, f_j)$ in the boundary-layer (figure~\ref{subfig:interaction_map}) between the fundamental and first-harmonic waves of the second-mode. The first-harmonic wave displays a short wavelength structure which amplitude peaks towards the end of the cone and then displays a slow damping as the boundary layer enters the expansion region at the cone-cylinder junction. Such damping is also visible on the interaction map $ \pmb \psi_{i,j}(x, y, f_i, f_j)$ of figure~\ref{subfig:interaction_map} where the self-interaction of the second-mode at $f_{2nd}=175$kHz peaks before the end of the cone and vanishes in the expansion region. The fast decay of the interaction coefficient $\pmb \psi_{i,j}$ after the cone-cylinder junction points toward a damping effect of the expansion on the triad interaction. This is consistent with the decay of second mode amplitude in this region, which in turn has less energy to transfer to its harmonic. The number of possible interactions observed and the clear evidence of second mode harmonics in the dataset highlight how far is the flow in the non-linear regime before the separation.

\subsection{Discussion on the scenario}
This section highlighted a case where the Reynolds number is high enough so that attached boundary layer dynamics leads to transition.
For the presented case, second Mack mode waves get amplified on the cone until they reach non-linear saturation.
Before the expansion fan, non-linear interactions have already begun to fill the spectrum, hinting that the flow is near breakdown to turbulence.
Then, even if the expansion fan damps the instabilities, breakdowns happen either in or upstream of the SBLI, leading to incipient separation. 
For higher Reynolds number, breakdown will continue to move upstream, still caused by second mode driven breakdown, and cause the absence of a separated region.

For lower Reynolds number, given that breakdown will be delayed, the separation bubble will grow in size (see \cite{marxenDisturbanceEvolutionMach2010,lugrinTransitionScenarioHypersonic2021,lugrin2022transitional}) and the presence of both a recirculation region and a mixing layer may lead to other transition scenarios, this will be studied in the next section.


\section{Transition with a large separated region on the cylinder and flare} \label{sec:large-sep}

This section aims at studying the transition scenario for regimes where two conditions are respected : breakdown happens on the geometry and a large separated region is present.
Owing to the flare spectra shown in figure~\ref{fig:PSD-re}, transition seems to first happen between $\Rey_\infty = 3.77 \times 10^6$ and $\Rey_\infty = 4.24 \times 10^6$, going back to figure \ref{fig:static-trends}, at that Reynolds number, the separation region is close to its largest size.
For those conditions, PCB05 to 10 are located inside the recirculation region.


\subsection{Streamwise evolution of wall-measurements at $\Rey_\infty=3.82\times 10^6$}

\begin{figure}
    \centering
    \includegraphics[width=\textwidth]{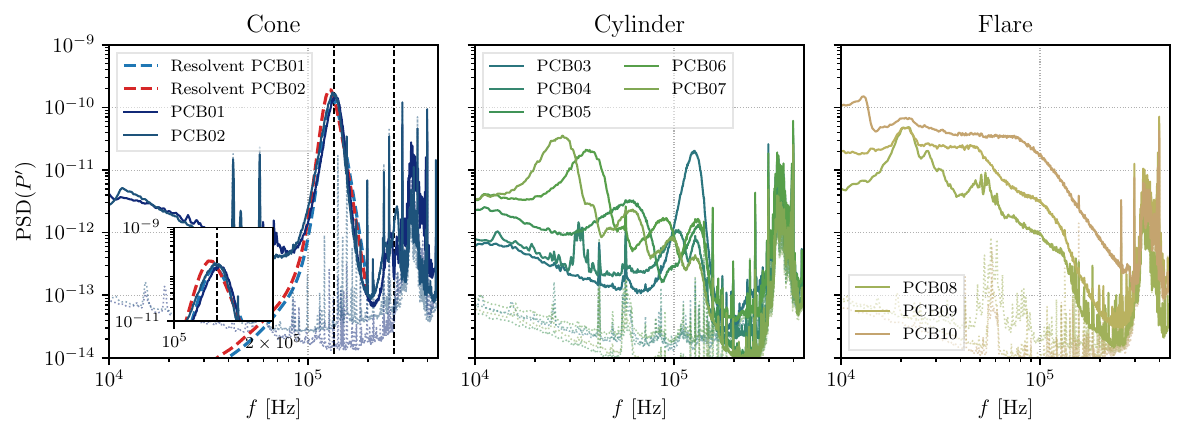}
    \caption{PCB wall pressure spectrum evolution along the geometry for $\Rey_\infty=3.82\times 10^6$. Comparison with reconstructed wall pressure from the Resolvent analysis scaled by an amplitude $A_0$ at the first PCB.}
    \label{fig:pcb-geom-5380}
\end{figure}

A first focus is set on the PCB spectra evolution at a selected Reynolds number of $\Rey_\infty=3.82 \times 10^6$, such an evolution is given in figure~\ref{fig:pcb-geom-5380}. 
The 10 PCB sensors placed along the object exhibit clear peaks at every stations along the geometry. 
On the cone, two second-mode peaks can be observed around $135$kHz. A slight shift toward lower frequencies and higher amplitudes is seen between PCB01 and PCB02. The shift toward lower frequencies for PCB02 is consistent with the thickening of the boundary layer and the observed growth without peak broadening suggests a linear behaviour of the second mode on the cone.

As it was observed for higher Reynolds number, at the end of the cone, as the second-mode passes through the  expansion fan, its amplitude decreases by one order of magnitude. 
Again, such behaviour is consistent with the previous work of \cite{butlerInteractionSecondmodeWave2021} on the dampening effect of expansion corners on second mode waves. 
Such a stabilisation of the second mode is captured on the downstream sensors up to the cylinder-flare junction. Indeed, for PCB03, the amplitude of the high-frequency peak has been reduced by one order of magnitude in comparison to the cone measurements. 

On the other hand, for lower frequency content, no signature is detected on the cone. However, an amplification of multiple peaks below 100kHz at successive stations is visible on the cylinder in figure~\ref{fig:pcb-geom-5380}b from PCB04 to PCB07, starting from 2 distinct peaks on PCB04 to up to 4 successive peaks on PCB07.
This amplification pursues along the flare as shown in figure~\ref{fig:pcb-geom-5380}c, where PCB08 and PCB09 display a strong wall-pressure peak for $f=20$~kHz.
For PCB10, spectral broadening is occurring as the flow is starting to display characteristics typical of turbulence.
A remarkable feature of this region is that there are numerous (up to 4) distinct peaks captured in the PSD shown figure~\ref{fig:pcb-geom-5380}b. 
This observation contrast with previous experimental observation describing only a low and a high frequency peak in the separated region \citep{benitezInstabilityTransitionOnset2023}, where such peaks were respectively identified as a shear-layer mechanism and a second-mode like mechanism.
It should be noted here that these previous studies also identified possible non-linear energy transfers between these latter modes through bicoherence analysis. Suggesting that the second mode directly interacts with the first mode in the transition scenario. 
We aim at clarifying these dynamics in the following subsections.

\subsection{Understanding the wall pressure signature in the separated region}

\begin{figure}
    \centering
    \includegraphics[width=\textwidth, trim = 20 30 20 30,clip]
        {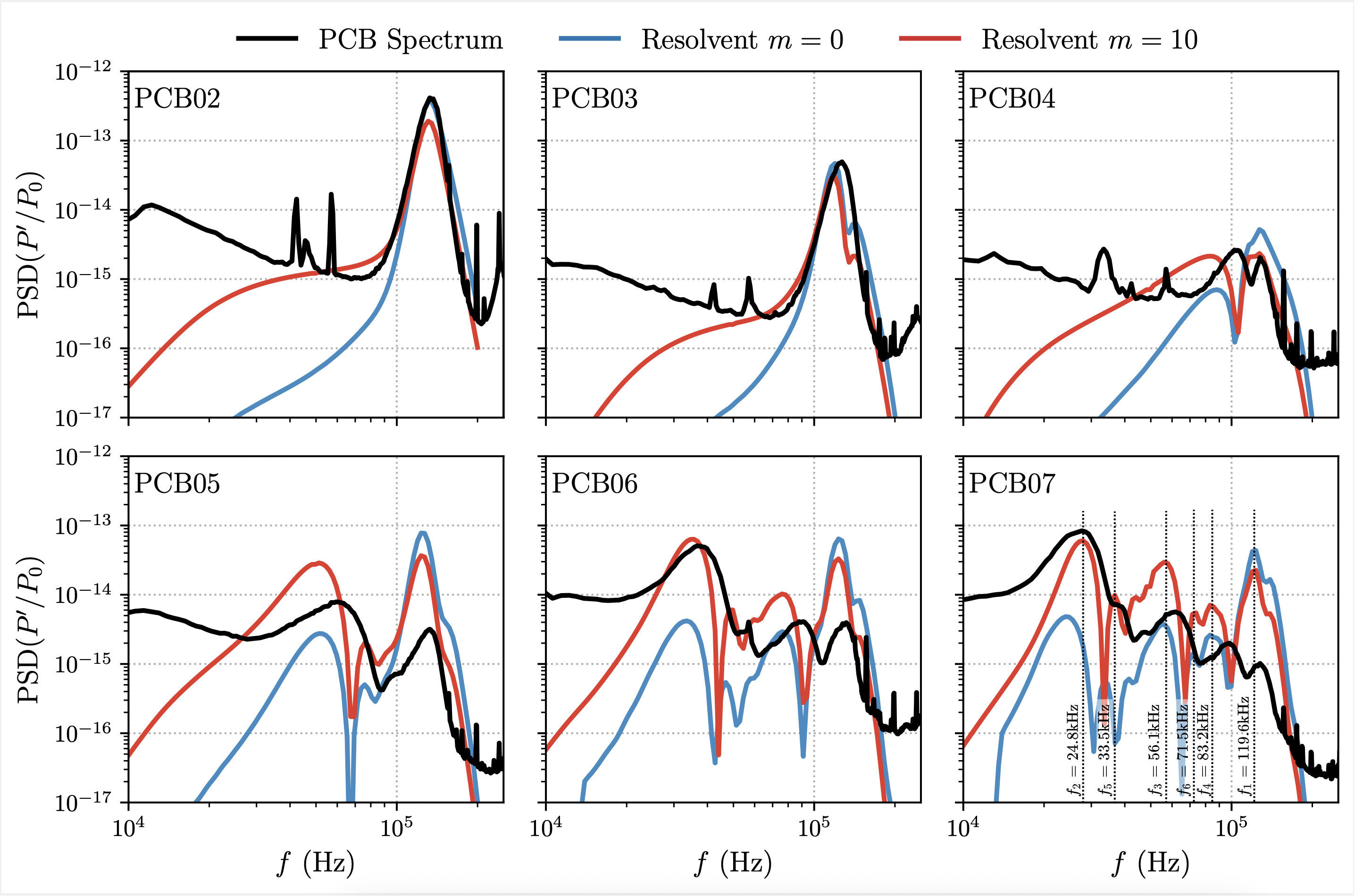}
    \caption{PCB wall pressure spectrum evolution along the geometry for $\Rey_\infty=3.82\times 10^6$. Comparison with Resolvent wall-pressure spectra for $m=0$ and $m=10$}
    \label{fig:pcbs_resolvent_case1}
\end{figure}

In order to get a more precise overview of the multiple pressure peaks in the separated region, figure~\ref{fig:pcbs_resolvent_case1} shows independent PSD plots of the successive PCB-sensors from the beginning of the cone to the flare. 
These sensors show the evolution of the wall pressure signature of boundary layer instabilities through the expansion fan and then the SBLI and separated region. 
To complement this experimental data and help to identify the convective mechanism underlying the peaks, PSD spectrum reconstructed from the optimal response of the global Resolvent analysis are added over the experimental PSD plots for $m=0$ and $m=10$. The pressure response of the Resolvent is only rescaled by a factor $A_0$ to match the amplitude of the second-mode peak at PCB01 and no other scaling is added as the optimal response evolves downstream. The same scaling $A_0$ is used for $m=0$ and $m=10$.

The Resolvent optimal gain map shown in figure~\ref{subfig:gain038} displays two peaks of amplification for unsteady waves on this baseflow. 
These peaks were previously identified as related to the first- and second-mode mechanisms over the geometry. However, the local perspective of figure \ref{fig:pcbs_resolvent_case1} shows that when plotting the wall-pressure signature of the optimal responses at various frequencies $f$ and wavenumbers $m$, the optimal responses lead to the amplification of multiple frequency-separated, linear mechanisms which are visible as peaks in the Resolvent pressure signature, up to 4 main peaks are visible at the PCB07 station for both experiments and resolvent analysis. 
The emergence of these peaks in the GSA and in the experiments is discussed in what follows. 

%

Looking at the evolution of the wall-pressure spectrum from the PCB02 in figure~\ref{fig:pcbs_resolvent_case1}a and as expected, the initial linear stages of the second mode growth are perfectly captured by the Resolvent analysis. The planar modes at $m=0$ are the one that align the best with the second mode peak amplitude. This indicates that the wall-pressure signature on the cone is most likely dominated by planar waves.
During these linear stages the second-mode exhibits growth on the cone, followed by a strong damping after the cone-cylinder expansion fan, visible on PCB03. 
At PCB04, the boundary layer is separated and both the Resolvent and experiments exhibit what seems to be a frequency splitting of the second mode peak.
These two peaks are noted to be at $f_1=127.0$kHz and $f_2=88.4$kHz.
Looking further downstream at PCB05, the frequency splitting further increases and the peak at $f_1$ shows a small frequency hump in the experiment and Resolvent around $f_3=73.7$kHz. 
At PCB06, the acoustic signatures at $f_1$, $f_2$ and $f_3$ are now clearly separated, underlining the presence of three distinct waves at this station. Another interesting finding lies in the match between the spectrum at $m=10$ and the experimental response for the low-frequency peak, suggesting a progressive switch in the dominant angle for low-frequency waves.
Finally, at PCB07, the mixing-layer is close to its maximum height and both the experiments and Resolvent display the presence of four peaks of frequencies $f_1$, $f_2$, $f_3$ and $f_4$, with the indice indicating the order of appearance of the peaks with the downstream evolution of the spectra.
It should be emphasized here, that the similar peak evolution between the Resolvent and the experiment means that these additional pressure peaks originate directly from the linear dynamics of the baseflow and are not the byproduct of non-linear interactions. This constitutes, to the knowledge of the authors, the first numerical and experimental confirmation of such dynamics in an hypersonic transitional separated flow.

It should be noted that discrepancies are observed between the Resolvent and experiments peaks amplitude in figure~\ref{fig:pcbs_resolvent_case1}.
These differences are attributed to the shrinking of the separated region induced by the breakdown at reattachment (see PCB08-10 of figure~\ref{fig:pcb-geom-5380}) and its non-linear impact on the separated region size. Additionally, we do not account for wind tunnel receptivity effects in the Resolvent analysis (see equation~\ref{eq:receptivity}) which may bias the real flow response towards specific frequencies. 
However, considering the good qualitative agreement between the experiments and Resolvent PSDs, it can be argued that these discrepancies remain relatively small and the focus is mainly set on the frequency peaks predictions. Therefore, the nature of the different observed peaks in the separated region will also be discussed using the optimal responses of the Resolvent on the laminar base flow.

\begin{figure}
    \centering
    \includegraphics[width=\textwidth]{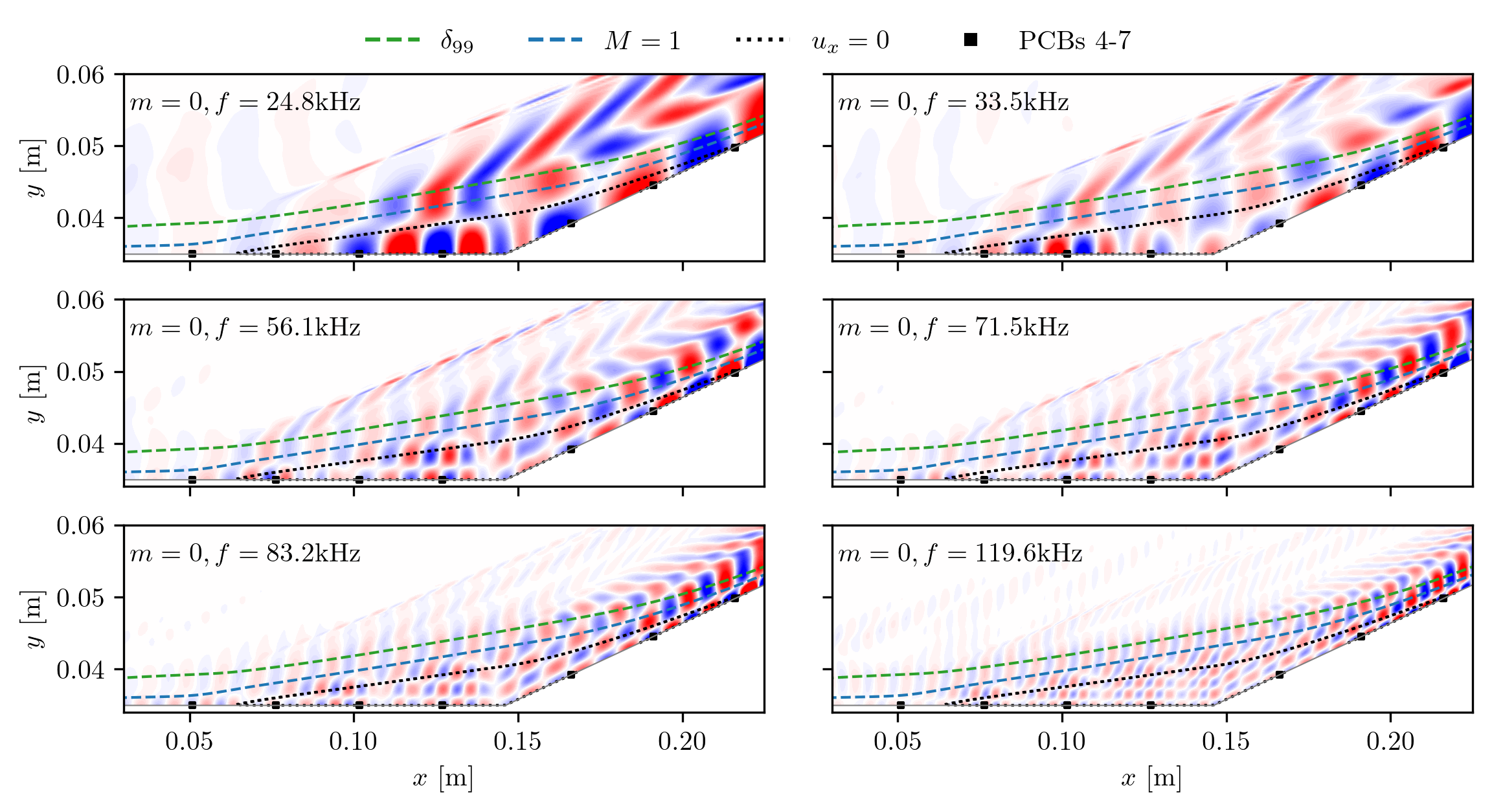}
    \caption{Normalised pressure contours of the optimal responses at the peaks identified from PCB07 spectrum}
    \label{fig:optimal_response_pcb07}
\end{figure}

The optimal responses at $m=0$ corresponding to each clearly visible peak of PCB07 on figure~\ref{fig:pcbs_resolvent_case1} are displayed in figure~\ref{fig:optimal_response_pcb07} with the main feature of the baseflow highlighted. 
The first visible difference between the modes is the decreasing size of the structures in the streamwise direction as the frequency increases. A second predominant feature is also visible in the separated region and lies in the number of wall-normal oscillations of the optimal response which also increases with the frequency. 
Each optimal mode shows a distinct pattern in the separated region and multiple zones of wall normal oscillations can be observed for a given optimal response (see modes at $f_4 and f_3$ for instance).
The optimal forcings for these modes, not shown here are all found to be located on the cone region, meaning that these bubble modes originate from a receptivity mechanism occurring on the cone. 
Finally, these wall normal oscillations seem to be mostly located below the sonic line ($M=1$), depicted with a blue dashed contour, akin to a second mode mechanism.

The increasing frequency gap between $f_1$ and $f_2$, the successive appearance of waves at $f_3$ and $f_4$ as the mixing layer height increases and their spatial organisation, invites us to conjecture a relation between the high and low frequency waves through the presence of trapped acoustic waves amplifying at selected frequencies within the separated region.
The following sections investigate two aspects of this hypothesis. First, the nature of the waves leading to the peaks $f_1$ to $f_4$ is verified using Resolvent and extraction of wall-normal pressure profiles.
Second, a geometric scaling is suggested to define the temporal scale of these waves. 


\subsection{Trapped acoustic waves in the bubble} \label{sec:trapped-waves}

\begin{figure}
    {\centering
     \begin{subfigure}{0.445\textwidth}
         \centering
         \includegraphics[width=\textwidth]{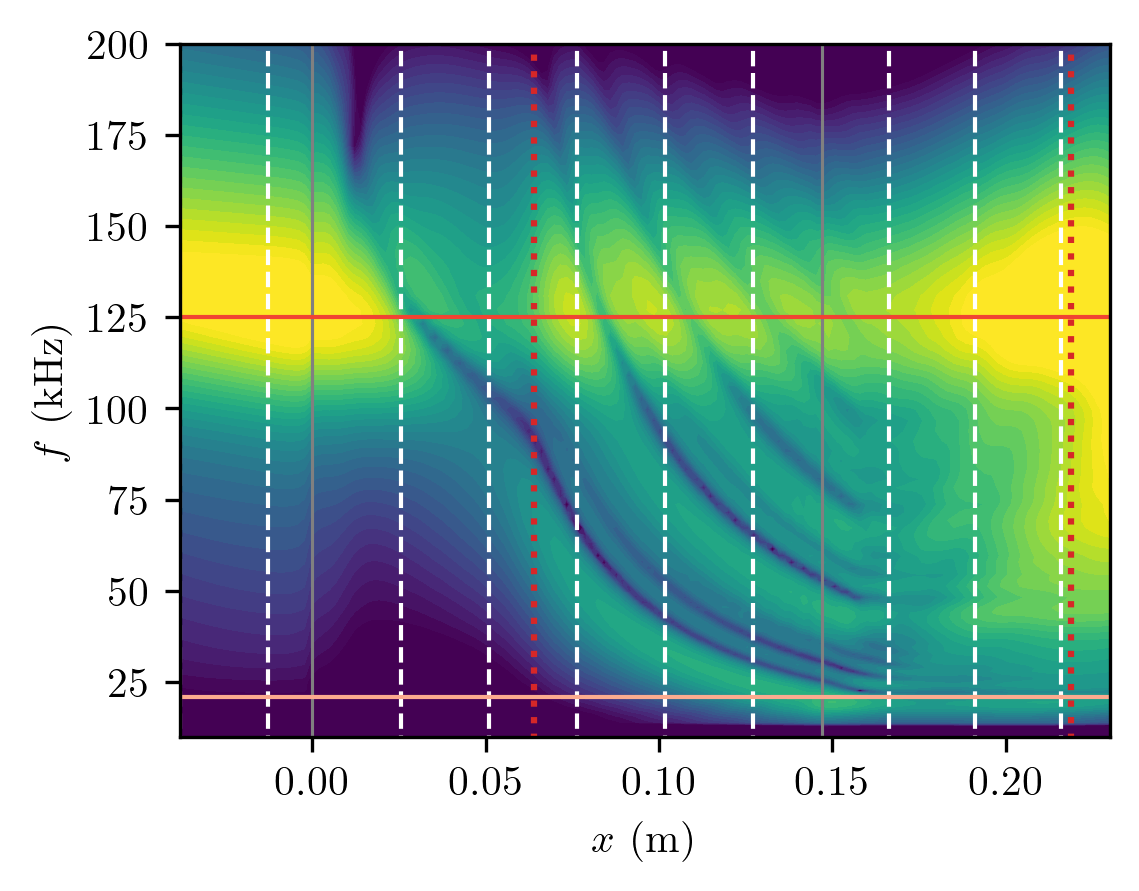}
         \caption{Frequency streamwise spectrum at $m=10$}
         \label{subfig:dim_re038_res0_spectrum}
     \end{subfigure}
     \hfill
     \begin{subfigure}{0.535\textwidth}
         \centering
         \includegraphics[width=\textwidth]{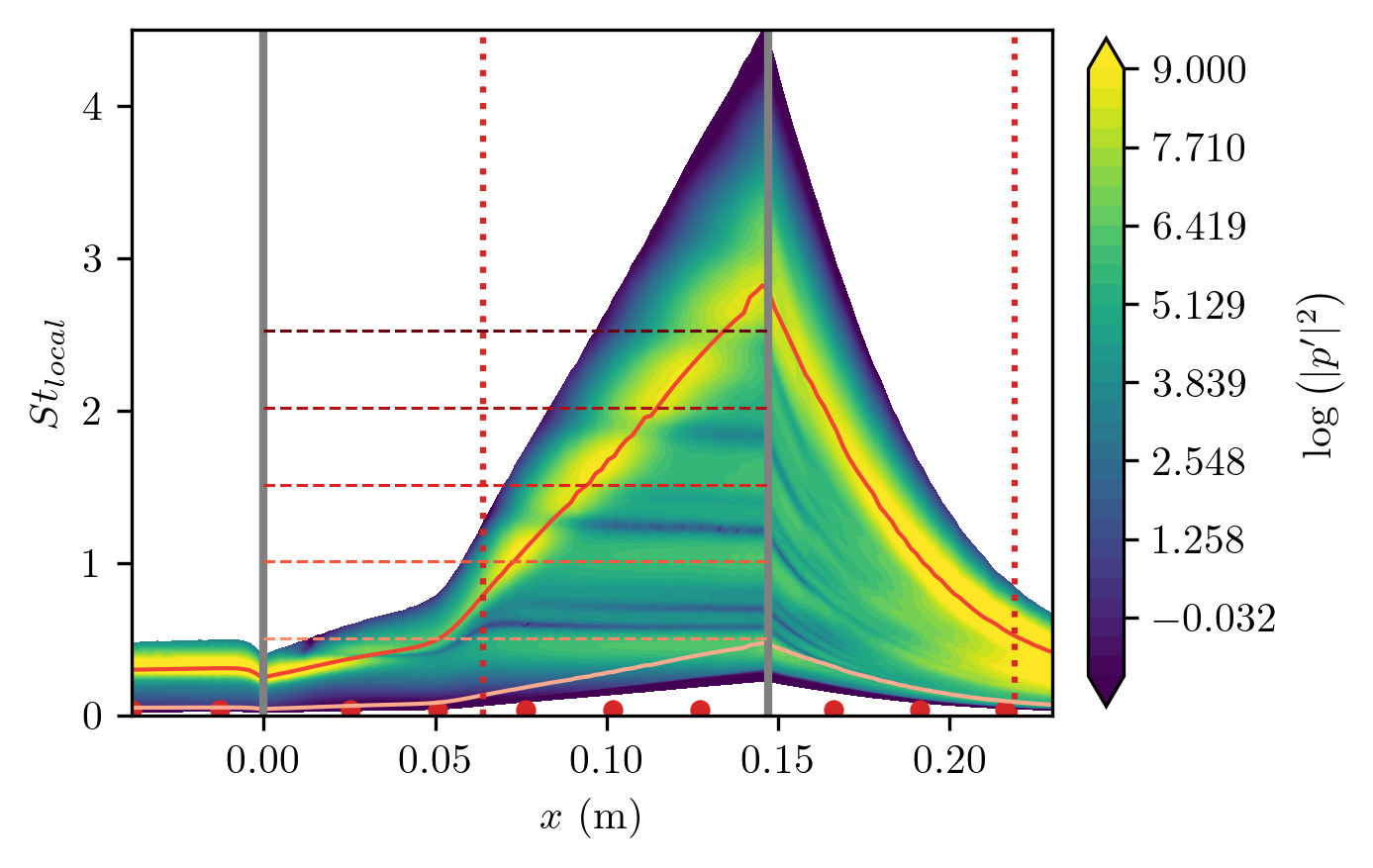}
         \caption{Scaled streamwise spectrum with $St_{\text{loc}}$ at $m=0$}
         \label{subfig:nodim_re038_res0_spectrum}
     \end{subfigure}
     \\
      \begin{subfigure}{0.445\textwidth}
         \centering
         \includegraphics[width=\textwidth]{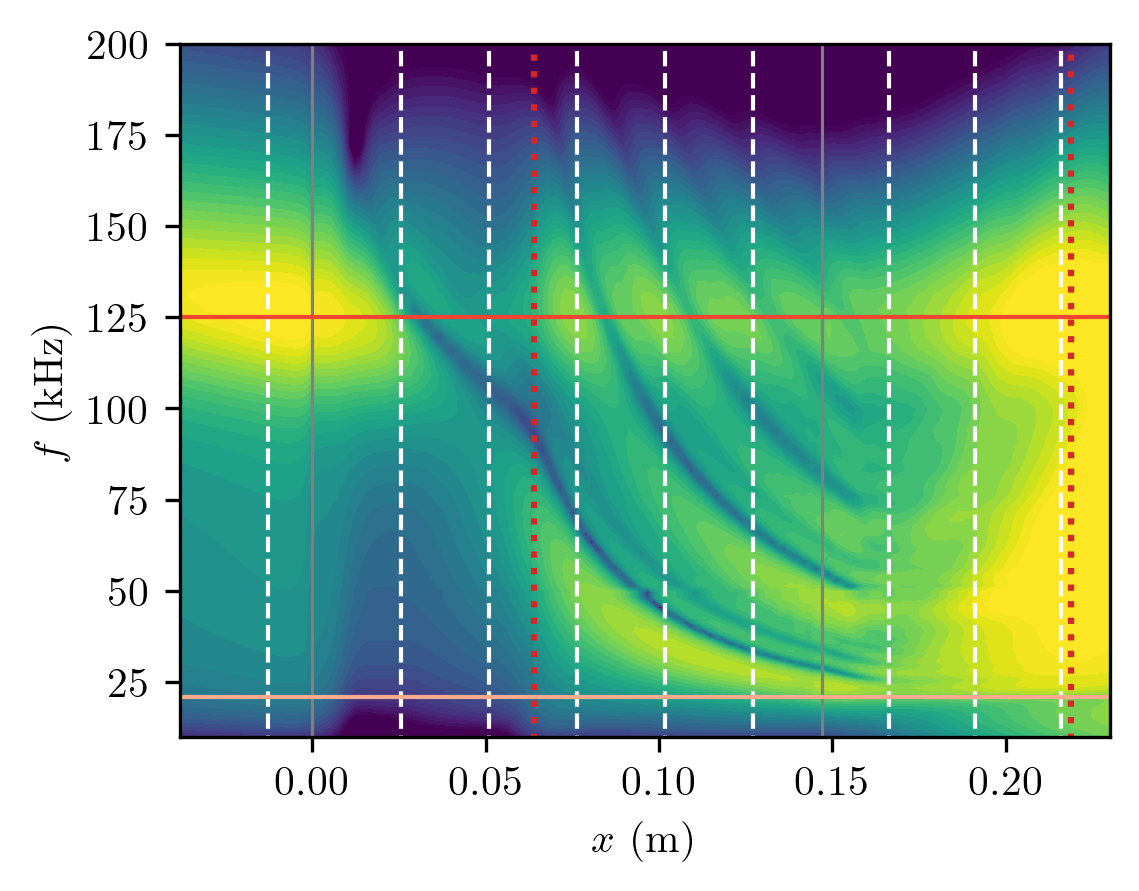}
         \caption{Frequency streamwise spectrum at $m=10$}
         \label{subfig:dim_re038_res10_spectrum}
     \end{subfigure}
     \hfill
     \begin{subfigure}{0.535\textwidth}
         \centering
         \includegraphics[width=\textwidth]{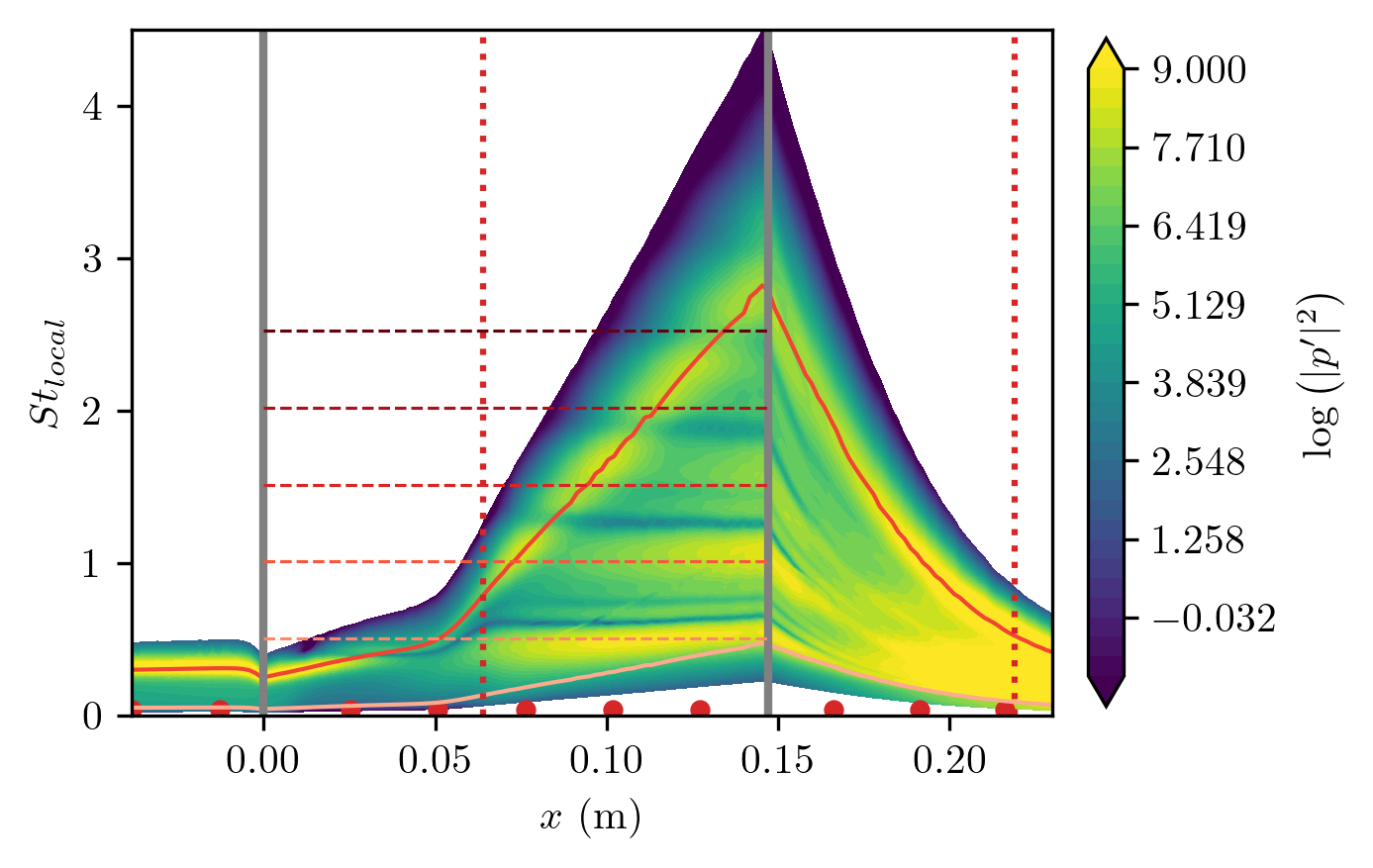}
         \caption{Scaled streamwise spectrum with $St_{\text{loc}}$ at $m=10$}
         \label{subfig:nodim_re038_res10_spectrum}
     \end{subfigure}
    \caption{Resolvent reconstruction of the wall pressure spectrum from the end of the cone to the beginning of the cylinder and flare. Dimensional and non-dimensional spectra are shown to illustrate the frequency selectivity of the separated region. PCB positions from PCB03 to PCB08 correspond to the white dashed lines (left column) or the red dots (right column). The separation is identified by the vertical red dotted lines. Second mode frequency is shown in a red continuous line, First-mode frequency is shown in a pink continuous line. }
    \label{fig:re038_res_spectrum}}
\end{figure}

Considering the good agreement between the Resolvent results and the PCBs measurements, the local spectra along the cylinder are complemented by the Resolvent analysis by computing a continuous wall pressure spectrum, in-between PCB measurements, from the end of the cone to the flare. This streamwise spectrum evolution is shown in figures~\ref{subfig:dim_re038_res0_spectrum} and \ref{subfig:dim_re038_res10_spectrum} for $m=0$ and $m=10$ respectively.
The previously discussed PCB spectra from PCB03 to PCB08 correspond to slices taken at the white dashed lines in this figure. The separation is identified by the vertical red dotted lines and the cone, cylinder and flare junctions are delineated by grey lines. 

Looking at the continuous wall spectrum, it clearly appears, that the flow exhibits three distinct regions of wall-pressure response.
A first region correspond to the part of the spectrum where a single wave peaking at $f=120$~kHz (continuous red line), corresponding to the second Mack mode, dominates. 
A second region localised on the flare displays a somewhat similar behaviour with a strong wall signature around $f=120$~kHz spreading towards lower frequencies.
The third region corresponds to what happens in between, on the cylinder and more importantly in the separated region where up to five downward-curved wall pressure signature are visible for both $m=0$ and $m=10$. These curved wall signatures display successive amplitude peaks at $f=120$~kHz that dampen and shift toward lower frequencies as the flow evolves downstream, to the cylinder-flare junction. The lowest frequency, highlighted with a pink continuous line indicates the frequency of the peak usually identified as a shear layer mode in the literature.
Such successive amplifying lobes were previously observed in the local stability analyses of \cite{esquieuFlowStabilityAnalysis2019}, however no interpretation was given at that time as these results were believed to be an artefact of the local analysis for non-parallel flows and were not observed in the experiments. 
%

Noticing the link between the lower frequencies shift and the increase of the shear layer height within the separated flow as seen in figure~\ref{fig:optimal_response_pcb07}, a streamwise scaling of the spectrum based on a local Strouhal number $St_{\text{loc}}$ defined by the sonic line height is proposed in figures~\ref{subfig:nodim_re038_res0_spectrum} and \ref{subfig:nodim_re038_res10_spectrum}. This Strouhal number is defined as,
\begin{align}
    St_{\text{loc}}(x) =  \frac{f \times y_{\text{sonic}}(x)}{u_x(x,y_{\text{sonic}})}, \quad \text{where } M(x,y_{\text{sonic}})=1. \nonumber
\end{align}
Other scaling, such as a boundary-layer height based Strouhal were tested, but did not offer an interpretation as clear as this sonic line height scaling. Now, observing the scaled data shown in figures~\ref{subfig:nodim_re038_res0_spectrum} and \ref{subfig:nodim_re038_res10_spectrum}, the wall spectrum clearly exhibits pressure signatures at discrete Strouhal numbers (highlighted with horizontal red dashed lines) successively amplified along the separated region. These Strouhal number values will be specified in the next sections.
%
Additionally, the scaled data can be related to the dimensional spectrum by looking at the aforementioned red and pink continuous lines illustrating the $120$~kHz and $25$~kHz waves loci in the scaled spectrum. From these lines, it can be concluded that the $120$~kHz wave jumps to peaks at increasing Strouhal numbers throughout the separated region, each of these peaks corresponding to a higher order wave at a constant $St_{\text{loc}}$ with respect to the sonic line height. On the other hand, it should be noted that the $25$~kHz wave, often termed "shear-layer mode" in the literature, is related to a low Strouhal wave at the end of the cylinder region, corresponding to the maximal shear layer height. Looking at its upstream origin, it can be seen that this low $St$ mechanism is initially supporting the $120$~kHz wave, corresponding to the second mode, at the beginning of the cylinder. This finding questions the nature of the low-frequency peaks usually identified as first-mode waves and suggests a direct link with second-mode waves instead. 
\begin{figure}
    \centering
    \includegraphics[width=0.9\textwidth]{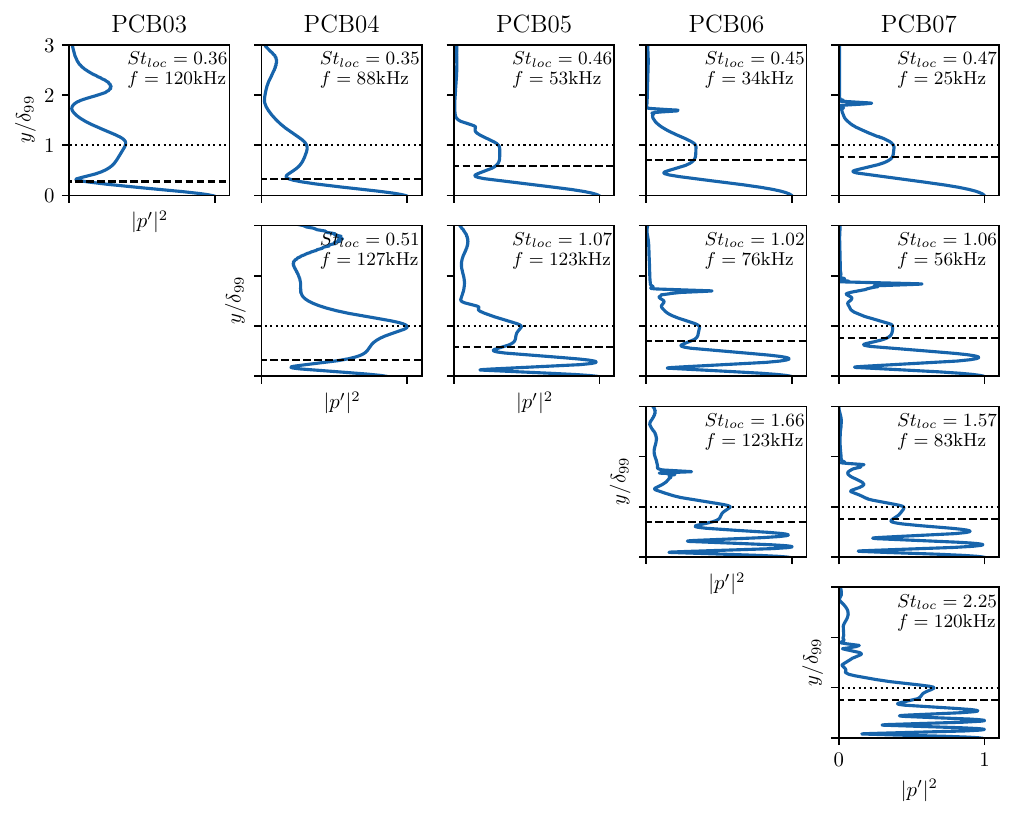}
    \caption{Streamwise evolution of the Resolvent optimal Response along the cylinder for highlighted peaks at $m=0$. Profiles at $m=10$ can be found in Appendix~\ref{annex:prof_m10}}
    \label{fig:case1_res_optimal_p_prof}
\end{figure}

Further evidence of the presence of such acoustic waves trapped below the sonic line is given in figure~\ref{fig:case1_res_optimal_p_prof}. Pressure profiles of the optimal Resolvent modes taken along wall-normal lines at the successive PCB stations located in the separated region are presented on each column, the locations are identified by red dots in figures~\ref{subfig:nodim_re038_res0_spectrum} and \ref{subfig:nodim_re038_res10_spectrum}. For each column, the successive rows correspond to increasing values of $St_{\text{loc}}$.
The values of $St_{\text{loc}}$ are chosen from the peaks of the frequency spectra shown in figure~\ref{fig:pcbs_resolvent_case1}
Therefore, figure~\ref{fig:case1_res_optimal_p_prof} displays the successive appearance of the different modes related to the increasing number of pressure peaks in the spectra. 
It appears clearly that each value of amplified $St_{\text{loc}}$ corresponds to a pressure wave having a given number of lobes below the sonic line.
\begin{table}
\centering
\begin{tabular}{lccccc}

 Index $n$ & $1$ & $2$ & $3$ & $4$ & $5$ \\
\hline
Frequency estimated (kHz) & $26.8$ & $53.6$ & $80.3$ & $107.1$ & $133.8$ \\
Strouhal estimated & $0.50$ & $1.01$ & $1.51$ & $2.02$ & $2.52$ \\
\hline
Frequency Resolvent (kHz) & $24.8$ & $56.1$ & $83.1$ & $119.6$ & $135.0$ \\
Strouhal Resolvent & $0.47$ & $1.04$ & $1.56$ & $2.26$ & $2.54$ \\
\hline
Relative error (\%) & $8.1$ & $4.5$ & $3.4$ & $10.5$ & $1.0$ \\
\hline
\end{tabular}

\caption{Comparing the frequencies and Strouhal number estimated from a duct acoustics analogy with the peak frequencies of the Resolvent wall-pressure spectrum at PCB07}
\label{tab:duct-analogy}
\end{table}
The number of lobes increases with the value of $St_{loc}$ as commonly found for trapped acoustic waves with a relation of the form $f = {n c_{s}}/{2L}$. With $n$ the wave order, $c_s$ the sound-speed and $L$ the characteristic length. An estimation of the frequencies and Strouhal numbers found from this approximation at the position of PCB07 is provided in table~\ref{tab:duct-analogy}. The sound-speed used for this estimation is taken as the mean sound-speed below the sonic line and the sonic line height $y_{sonic}$ is used for the parameter $L$, using the approximation of an infinite acoustic impedance at both ends. The estimated frequency values are closely matching the resolvent peak frequencies visible on figure~\ref{fig:pcbs_resolvent_case1}, with a mismatch of less than 10\% for each indice, except $n=4$, which might be blurred by the separation of two peaks at this PCB07 station. This analogy can then be useful to estimate the peak frequencies at the wall for such instabilities in a hypersonic separated region, where the fundamental peak can be found at $St\approx0.5$. Finally, it should be noticed that these trapped waves are reminiscent of the so-called Mack's higher order modes \citep{mackBoundaryLayerLinearStability1984a,fedorovHighSpeedBoundaryLayerInstability2011} and the possibility of their presence in such flow conditions was hypothetised recently \citep{caillaud2025separation} from a local analysis.
These above findings actually confirm the presence of such waves and relate them to the experimentally measured flow structures while offering a quick method to estimate their frequencies.

Another general finding related to this analysis lies in the nature of the measured low frequency wave, here found at $f=25$~kHz for PCB07. Previous literature identified this lower frequency peak at the wall, as being related to a shear-layer mechanism \citep{benitezInstabilityTransitionOnset2023,esquieuFlowStabilityAnalysis2019}. 
We show in figures~\ref{fig:pcbs_resolvent_case1}~and~\ref{fig:case1_res_optimal_p_prof} that the PCB-measured low frequency peak actually corresponds to a continuation of the second-mode wave as a trapped acoustic wave of low order (i.e. with $n=1$ oscillation under the sonic line), found for $St_{loc}\approx 0.47$. 
This can be seen in the first row of figure~\ref{fig:case1_res_optimal_p_prof}. This continuation is also observable in figures~\ref{subfig:nodim_re038_res0_spectrum} and \ref{subfig:nodim_re038_res10_spectrum} where at $x=0.25$m, the low $St_{loc}$ peak splits from the second mode peak at boundary layer separation, where the high-frequency peak (highlighted in red) follows higher-order trapped waves and the low frequency peak (highlighted in pink) ends up corresponding to a first-order acoustic resonance.
The same frequency splitting is also visible on figure~\ref{fig:pcbs_resolvent_case1} for PCB03,04 and 05 and such measurements can now be interpreted as a frequency separation of the different trapped waves, with the low order waves shifting towards lower-frequencies. This interpretation of the low-frequency pressure signature at the wall as being unrelated to first-mode waves is supported by three additional observations. First, the first-mode waves are known to only display a weak wall pressure signature \citep{bugeat3DGlobalOptimal2019}. Second, the cold wall conditions studied here are not favourable for first-mode waves growth \citep{mackBoundaryLayerLinearStability1984a} and finally, figure~\ref{subfig:chu_nrj_modes} shows that first-mode presents at least one-order of magnitude less energy in the cylinder region compared to second mode waves.

\subsection{Optical measurements in the separated region}
\begin{figure}
    \begin{subfigure}[b]{0.525\textwidth}
        \centering
        \includegraphics[width=\textwidth]{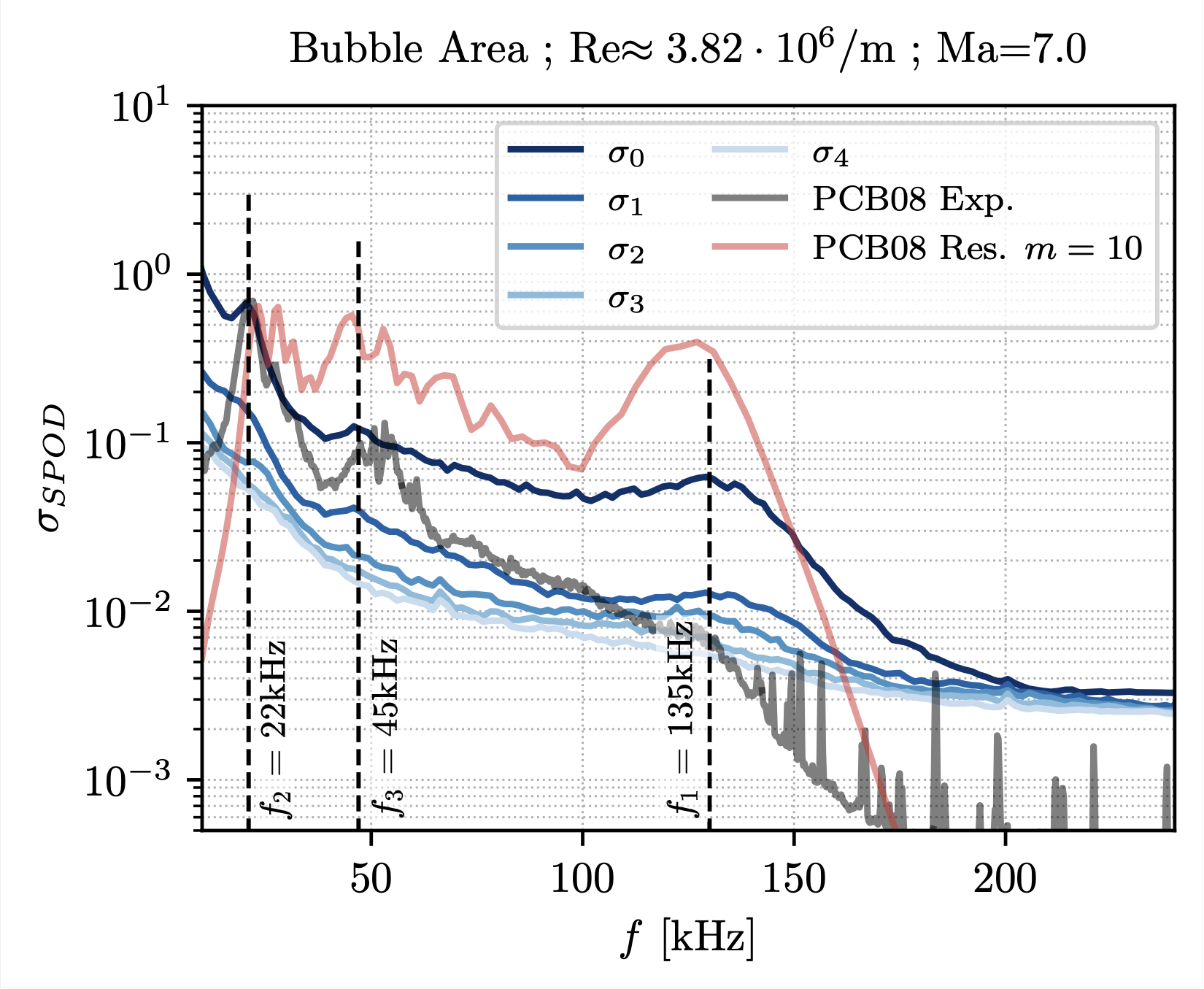}
        \caption{SPOD energy spectrum for the five first singular values $\sigma_0,...,\sigma_4$, compared to PCB08 on the flare}
        \label{fig:spod_nrj_5480}
    \end{subfigure}
    \hfill
    \begin{subfigure}[b]{0.47\textwidth}
        \centering
        \includegraphics[width=\textwidth]{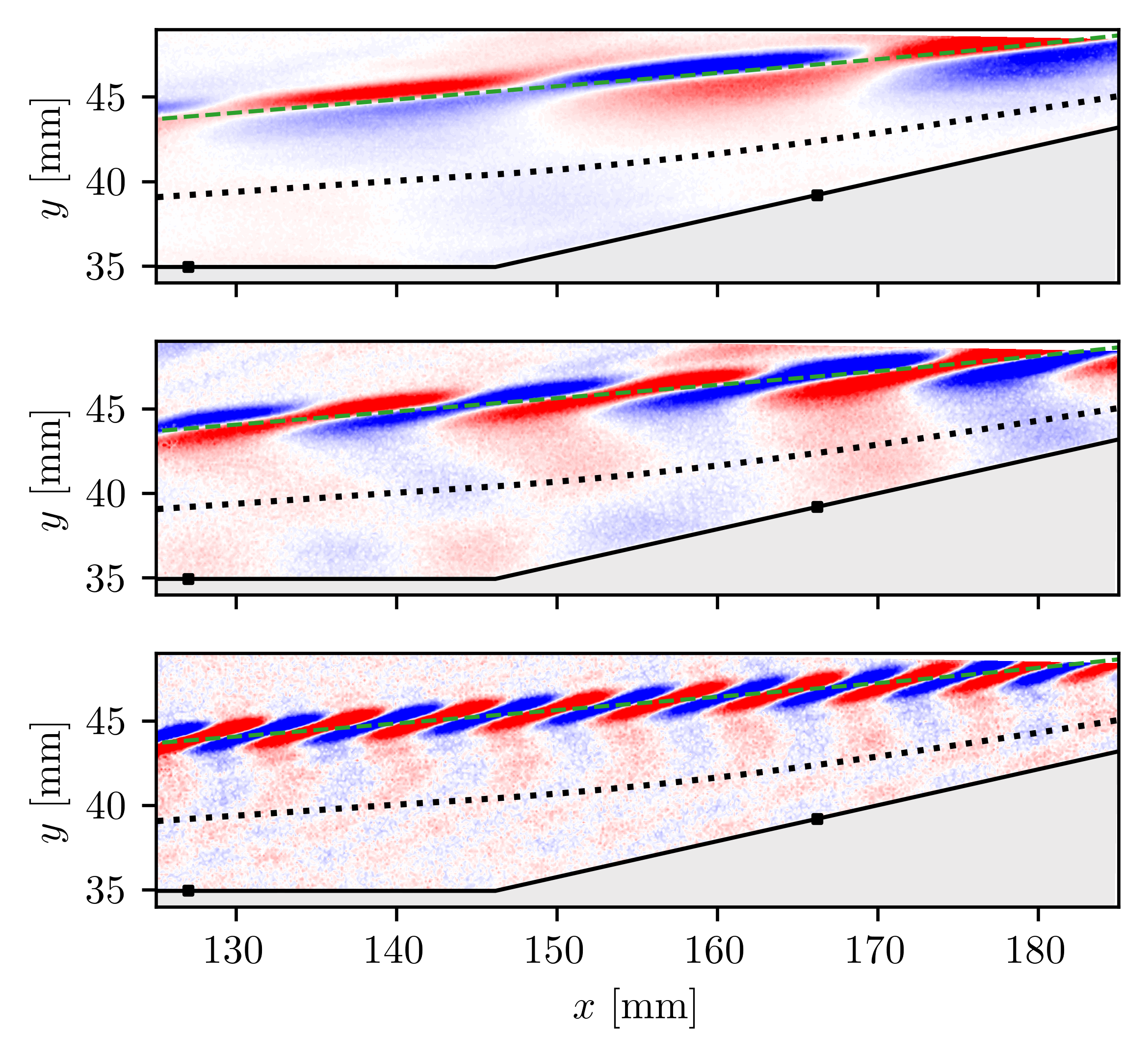}
        \caption{SPOD modes at the main energy peaks of (\ref{fig:spod_nrj_5480})}
        \label{subfig:spod_modes_5480}  
    \end{subfigure}
    \begin{subfigure}{\textwidth}
        \centering
        \includegraphics[width=\textwidth]{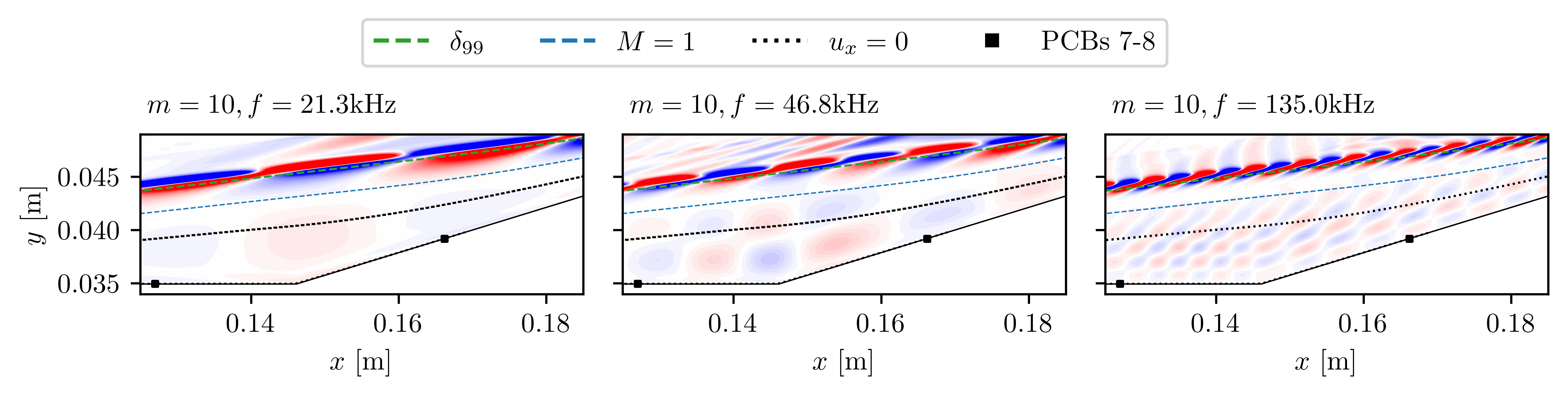}
        \caption{Numerical schlieren modes from the resolvent at the peaks of (\ref{fig:spod_nrj_5480})}
        \label{subfig:mean_flow_compare_038}
    \end{subfigure}
    \caption{SPOD analysis of the separated flow region at $\Rey_\infty=3.82\times 10^6$ with data from the PCB on the Flare. Informations from the numerical baseflow boundary-layer and recirculation region are added over the SPOD modes to help visualising the flow topology.}
    \label{fig:spod_modes_5480_all}  
\end{figure}
In figure~\ref{fig:spod_modes_5480_all}, the wall-pressure measurements are complemented by high-speed schlieren focused on the separated region. 
A SPOD analysis is performed on the unsteady schlieren dataset around the mean-flow. The SPOD energy spectrum and associated modes are shown in figures~\ref{fig:spod_nrj_5480} and \ref{subfig:spod_modes_5480}. For the latter, the numerical boundary layer height $\delta_{99}$ and $\bar{u}<0$ lines are shown to highlight the shear-layer and the recirculation region extent. A close match is found, confirming the experimental mean-flow to be close to the base flow used for the Resolvent analysis. 

The energy spectrum is found to display three main regions of amplification. First, a low frequency peak at $f_1=22$ kHz, corresponding to elongated density fluctuations in the shear layer (Figure~\ref{subfig:spod_modes_5480}) and large modes with only $n=1$ half period in the wall normal direction. 
Second, a high frequency peak at $f_2=135$ kHz, corresponding to short streamwise fluctuations peaking in the shear layer and an organised structure of weak amplitude in the recirculation region which matches the spatial structure of the same mode on figure~\ref{fig:optimal_response_pcb07} with $n=4$ half-periods in the wall normal direction. 
Finally, an intermediate peak at $f_3 = 45$~kHz shows a similar structure to the mode at $f_2$ but with longer wavelengths and larger structures in the recirculation region, having $n=2$ half-period of oscillation in the wall-normal direction. 

The SPOD energy spectrum in figure~\ref{fig:spod_nrj_5480} is also compared against the wall-measurement of PCB08 and its Resolvent counterpart. A close match of the dominating peak frequencies for $f_2$ and $f_3$ is found indicating that the SPOD data mainly reflects what happens at this streamwise location. For the peak at $f_1$, a moderate bump is visible on the SPOD energy and a large gap of amplitude is observed between the numerical and experimental PCB spectra indicating that non-linear saturation might be occurring for these frequencies at this station. This will be investigated in the next section. A direct comparison of the SPOD modes and Resolvent optimal responses in $\nabla \rho '$ is found in figures~\ref{subfig:spod_modes_5480} and \ref{subfig:mean_flow_compare_038}. The modes display a very similar spatial organisation, especially below the $M=1$ line, where the number of peaks and position of the density gradients peak are found to be really close.

These observations imply that the spatial structure of the Resolvent optimal responses can also be observed on the density gradient of the experimental disturbance field, further confirming the existence of the trapped waves in experimental conditions. It should be precised here that the density signature of the trapped waves is found to be much weaker than their pressure signature with the resolvent analysis as highlighted in figure~\ref{subfig:mean_flow_compare_038}, explaining why they remain weakly visible with the SPOD. Having identified the successively amplifying acoustic waves within the bubble, their potential role in the   experimentally observed non-linear processes is assessed in the next section.

\subsection{Discussion on the non-linear energy transfers} 

\begin{figure}
    \centering
    \begin{subfigure}{\textwidth}
        \includegraphics[width=0.9\linewidth]{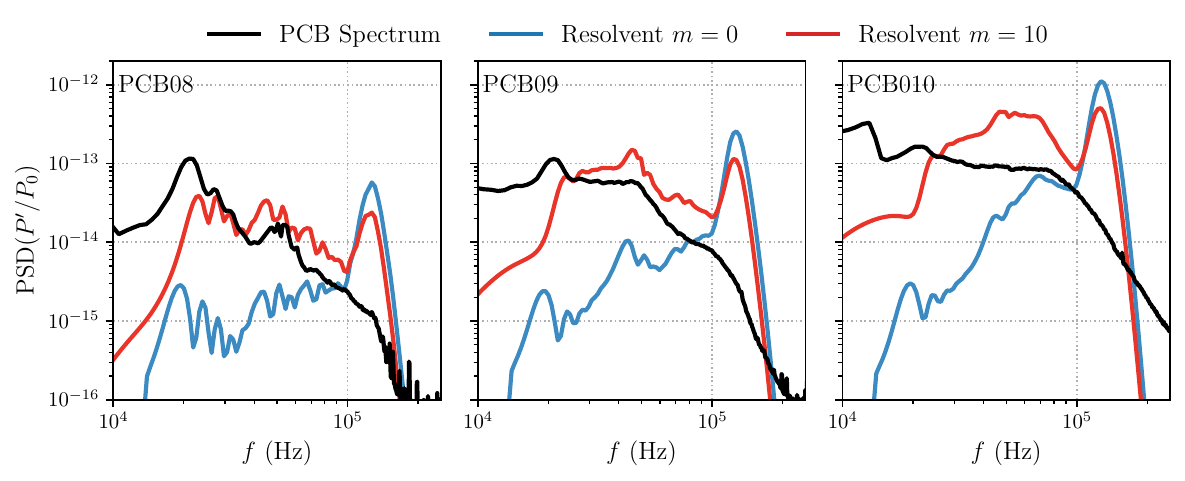}
        \caption{PCB wall pressure spectrum evolution along the flare for $\Rey_\infty=3.82\times 10^6$. Comparison with Resolvent wall-pressure spectra for $m=0$ and $m=10$, using the same $A_0$ amplitude defined before.}
        \label{subfig:res_expe_last_pcbs}%
    \end{subfigure}
    \\
    \begin{subfigure}{\textwidth}
        \centering
        \includegraphics[width=\linewidth]{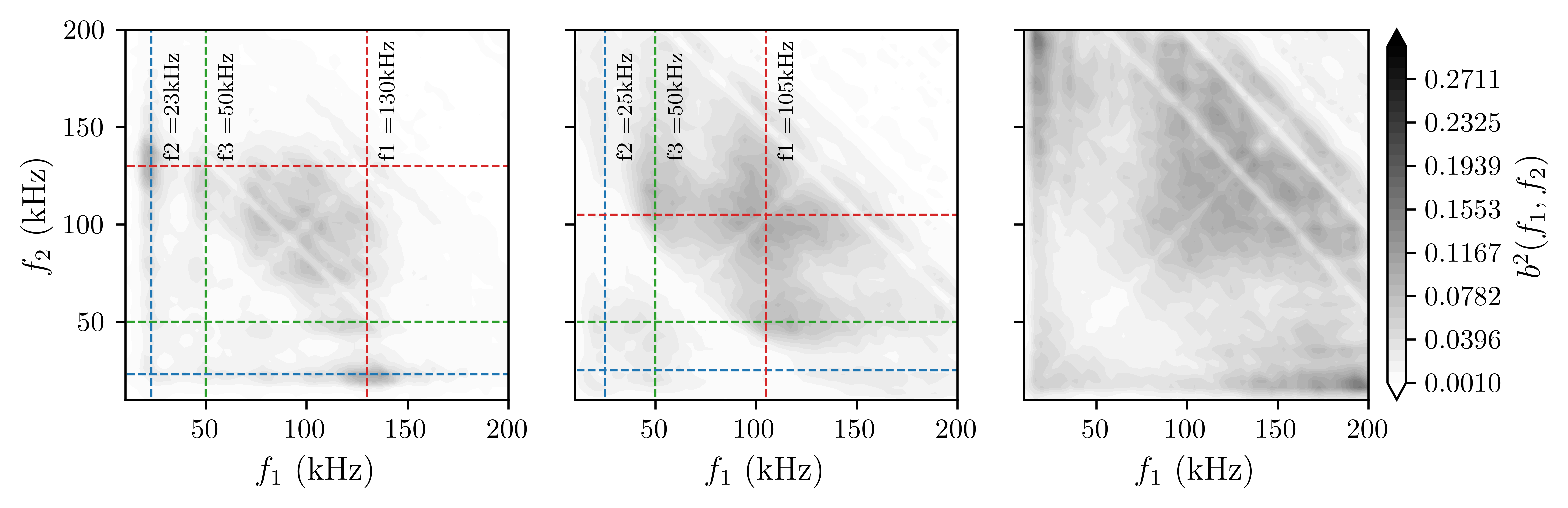}
        \caption{Bicoherence estimates from the PCB 8, 9 and 10 along the flare}
        \label{subfig:flare_bicoherence}%
    \end{subfigure}
    \caption{Linear and experimental summary of the breakdown in the flare region for PCB08-10}
    \label{fig:flare_breakdown}
\end{figure}

A bicoherence analysis is performed for the last three PCB sensors along the cylinder and the flare PCB sensors. This analysis aims at providing further insights in the previously discussed non-linear interactions between low-frequency and high-frequency waves across the separated region \citep{benitezInstabilityTransitionOnset2023}. The figures~\ref{fig:pcbs_resolvent_case1} and \ref{subfig:res_expe_last_pcbs} show that the experimental and numerical estimations start to significantly deviate from each other in the higher frequencies between PCBs 5 and 7. In the experiments the high-frequency peak is progressively damped, while the linear optimal response displays a continued amplification of this same peak.
In the meantime, the amplitude of the low-frequency mode at $f_2$ ($St\approx0.5$) is increasing up to PCB 9. Having higher amplitude than the Resolvent optimal responses, for which the low-frequency peak progressively vanishes. This is particularly visible for PCBs 8 to 10 and the SPOD (figure~\ref{fig:spod_nrj_5480}) where the low frequency peak dominates and the higher frequency content becomes broadband. In what follows, the  evolution of the measured spectra is investigated using bicoherence estimates.

\begin{figure}
    \centering
    \includegraphics[width=\linewidth]{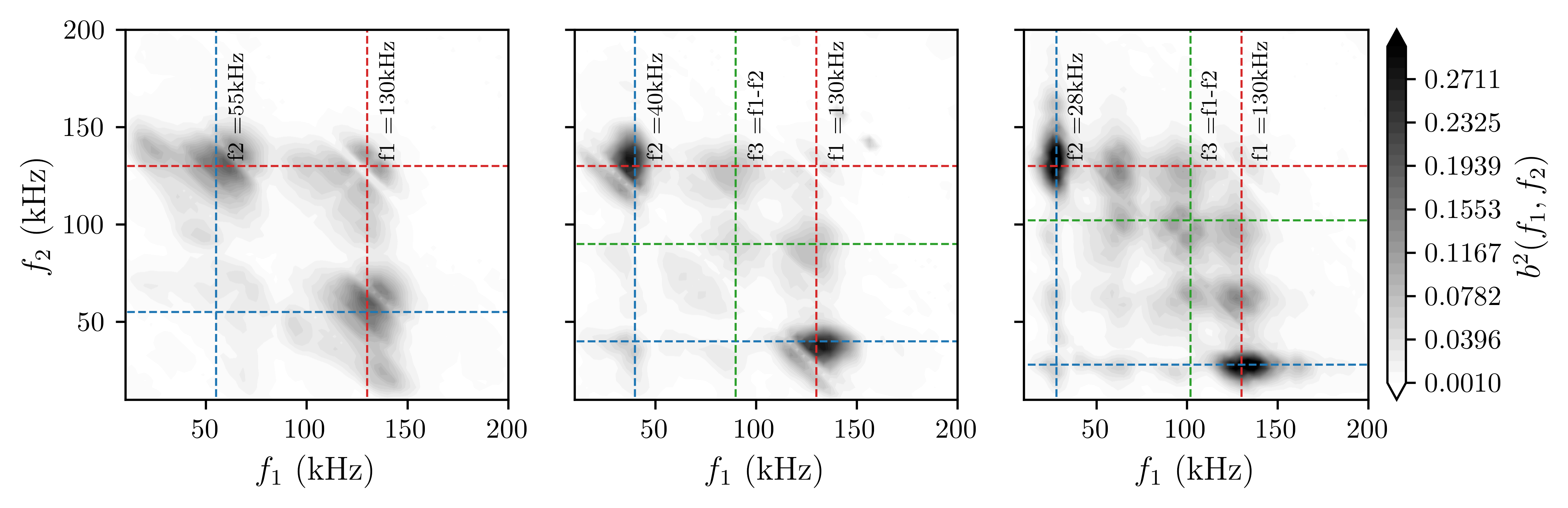}
    \caption{Bicoherence of sensors PCB05, PCB06 and PCB07}
    \label{fig:bicoherence_re38}
\end{figure}

 The sensors on the cylinder measured the multiple peaks of the trapped acoustic waves shown in figure~\ref{fig:pcbs_resolvent_case1}. The bicoherence maps of figure~\ref{fig:bicoherence_re38} show evidence of triadic interactions in the form of $f_1 \pm f_2 = f_3$ between the linear instabilities. 
In details, for PCB05 on figure~\ref{fig:bicoherence_re38} the bicoherence peaks correspond to a self interaction of the second mode at $f_2$ with itself in conjunction of an interaction of the low-frequency mode at $f_1$ with $f_2$. Further downstream the number of interactions increases while the self-interaction of the second-mode vanishes, the energy transfer between low and high-frequencies remains and a third phase-locking is found between $f_1$ and $f_2$ along with a self interaction of $f_2$. Going to PCB07 on figure~\ref{fig:bicoherence_re38}, these interactions are still visible, and a last phase locking is added between $f_3$, $f_4$ and $f_2$ generating multiple additional self and cross-interactions. 

The number of bicoherences peaks, in the upper half diagonal, grows from 2 at PCB05 to 8 at PCB07 as the flow evolves downstream. Notably, a possible dominant energy transfer mechanism is observed between the low-frequency wave at $f_2$ and the high frequency wave at $f_1$ throughout the bicoherence plots. In previous literature, this phase locking has been identified as an interaction of the shear layer-mechanism with the second mode \citep{benitezInstabilityTransitionOnset2023}. However, with the hindsight of the acoustic modes analysis discussed in section \ref{sec:trapped-waves} the phase-locking measured at the wall by the PCBs is found to be directly related to the trapped acoustic waves of different orders in the subsonic region. This suggests the possibility of a progressive transfer of energy from the leading second-mode peak of PCB 4 at $m=0$ to the low order $(S_t\approx 0.5)$ acoustic wave at $m=10$, found dominant for the PCBs 8 and 9 before breakdown, through the amplifying trapped acoustic waves in the subsonic region.

However, clearly establishing a non-linear energy transfer with the limited observations from the experiments remains difficult. Recalling that the bicoherence is a statistical estimator, it only provides information about possible phase-correlations of signals at different frequencies matching triad relation, but it cannot offer conclusions about causal energy transfers (equation~\ref{eq:bicoherence}, \cite{bountinEvolutionNonlinearProcesses2008}). Hence, at the difference of the cone transition scenario at $\Rey=6.06\times10^6$ shown in Figs.~\ref{fig:case5469_bicoherence_analysis} and \ref{subfig:cone-spod-nrj} where a second mode harmonic matching the bicoherence peak was found in the SPOD energy spectrum of the experimental data. The current experimental and numerical evidence provided for the $\Rey=3.82\times10^6$ case only display peaks between $f_1$ and $f_2$ (i.e. low and high order acoustic waves), making it difficult to establish whether a simple phase-correlation between the trapped acoustic waves at is observed at these bicoherence peaks or an actual causal energy transfer between the low and high order trapped waves is occurring. Indeed, no clear evidence of peaks originating from non-linear energy transfers such as harmonics are found and most of the peaks are described by the linear theory. This constitutes in itself an important conclusion regarding the analysis of experimental bicoherence peaks in a transitional separated region. 

Nonetheless, non-linear energy transfers have been seen to act as a forcing to linear mechanism, such as in oblique breakdown \citep{lugrinTransitionScenarioHypersonic2021}. Supporting this idea, it can be noticed in figure~\ref{subfig:res_expe_last_pcbs} that the experimental PSD spectrum progressively deviates in amplitude from the Resolvent optimal pressure spectrum. Especially between PCB08 and PCB10 where the experimental spectrum is clearly becoming broadband and deviating from the numerical data. This is further confirmed in figure~\ref{subfig:flare_bicoherence}, where interaction peaks can be observed at PCB08 and a broadband phase correlation, typical of the last non-linear stages of transition, is found for PCB10. This growing region of interactions above 100kHz is visible for the bicoherence plots of PCBs 7 to 10. Since no super-harmonics of the peaks between 90-150kHz are found in the wall measurements or the SPOD energy (figure~\ref{fig:spod_nrj_5480}), we suggest the presence of subtractive interaction in the form of $f_i - f_j = f_k$. Suggesting that the energy is progressively exchanged from high- to low-frequencies using intermediate frequencies at $f_3$ and $f_4$. 

Finally, this section presented the first experimental observations of trapped acoustic waves, which are confirmed by linear stability analysis. With these findings, the non-linear dynamics of the separated regions are observed with a new perspective. However, the transition scenario in this case remains to be completely elucidated since we cannot rule out the presence of an oblique breakdown as previous studies discussed it in those flows \cite{lugrinTransitionScenarioHypersonic2021,caoUnsteadyEffectsHypersonic2021}. Such questions will be answered with future numerical investigations where causal energy transfers could be measured between the different coherent structures corresponding to these trapped waves in the separated region.

\section{Discussion} \label{sec:discussion}

This study presents a detailed investigation of linear and non-linear dynamics in transitional hypersonic boundary layers developing over a Cone-Cylinder-Flare configuration, with important implications for the understanding of transition pathways of separated flow. The work provides new insights into how transition scenarios evolve with Reynolds number, demonstrating that fundamentally different mechanisms can govern the transition process under varying flow conditions.

\begin{figure}
    \centering
    \includegraphics[width=\textwidth]{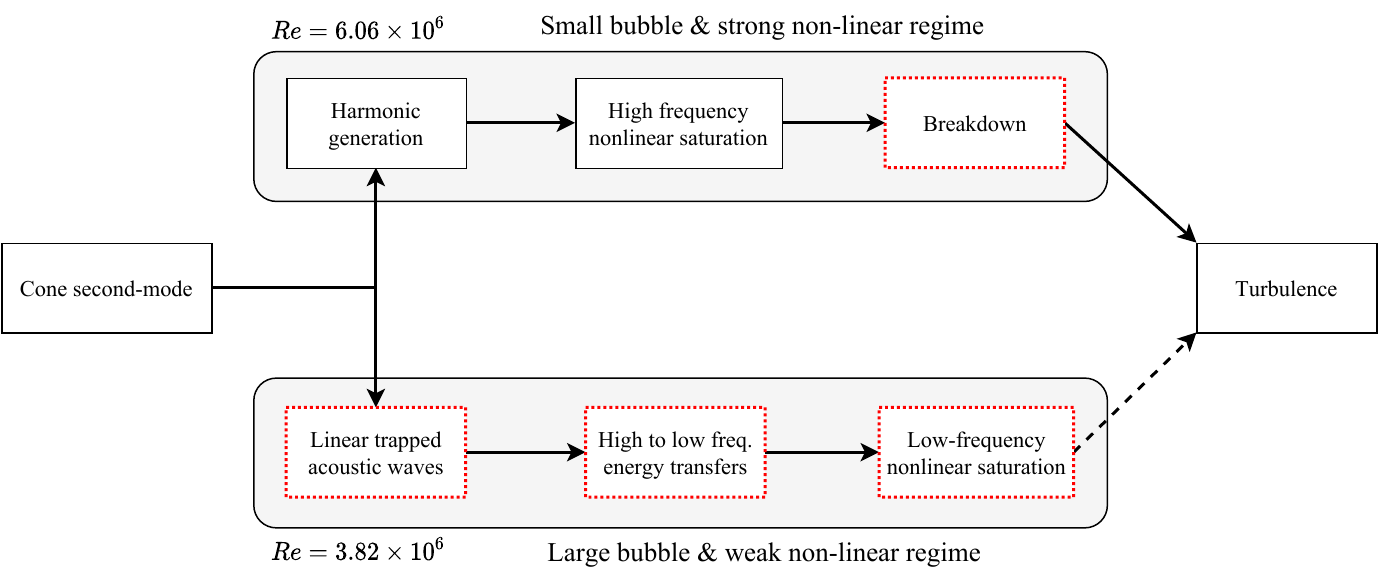}
    \caption{Summary of the steps to transitions for the two regimes observed in the experiments and computations. Steps happening in the separated regions are highlighted in red.}
    \label{fig:scenario_summary}
\end{figure}

A combination of numerical tools such as global linear stability analysis using the Resolvent operator and comprehensive experimental diagnostics such as PCB, IC2 sensors or SPOD and BMD of the high-speed schlieren were used to assess the flow dynamics measured in the wind tunnel. In accordance with past research, the transitional mean-flow observed in the experiments shows a separation length reduction as the Reynolds number increases. While, for its numerical laminar counterpart the separated region keeps extending, pointing towards a direct impact of the transition process on the separation length. Considering the coupled instability and separation length dynamics, the study focused on the nature of the linear mechanisms driving the separated region and its subsequent transition.
Two main scenarios were discussed,  a first route to turbulence, corresponding to incipient separated region at a Reynolds number of $\Rey=6.06\times 10^6$ and higher. This regime is dominated by second mode waves growing on the cone and then interacting non-linearly until breakdown. Then, a second route, at a lower Reynolds number of $\Rey=3.82\times 10^6$, where the separated region remains large and close to the fully laminar regime. In this case, complex waves growth and energy exchange were found in the measurements and described for the first time in the separated region. A summary of the steps related to these two scenarios is shown in figure~\ref{fig:scenario_summary}.

For the first case, the second mode evolution from the cone to the cylinder was studied in detail as it sets the linear and non-linear structures entering the separated region. At this Reynolds number the flow is found to transition quickly at the cylinder flare junction. On the flare, broadband wall pressure spectra are observed, indicating the end of transition and turbulence. At the initial step of transition on the cone, non-linear interactions were observed in the wall pressure sensors, with up to four harmonics between 160kHz and 800kHz, indicating that energy transfers between different timescales are already happening. The non-linearly saturating second mode on the cone-flare junction was also extracted from the time resolved schlieren SPOD with a level of detail not found in the literature before. The obtained experimental SPOD mode is shown to align well with the numerical reconstruction from the Resolvent optimal response, displaying growth on the cone and damping on the cylinder after radiating acoustic energy outward at the expansion fan. Using the schlieren dataset, further information on the non-linear interactions occurring higher in the boundary-layer are provided by performing Bispectral Mode Decomposition. This non-linear orthogonal decomposition allows us to observe the production of the first harmonic of the second-mode by triad interaction. Using the BMD information, the responsible flow region for the phase coupling between the fundamental second-mode and this harmonic in the boundary layer between the cone and the expansion are shown for the first time.

A second route with a large separated region at a lower Reynolds number of  $\Rey=3.82\times 10^6$ is then discussed. For this case a broadband spectrum is observed only at the end of the flare, indicating a late transition. The comparison of the linear analysis and experiments indicated that the incoming instabilities from the cone and before the separated region remain close to the linear regime. Further analysis of the wall-pressure sensors on the cylinder shows multiple energy peaks at different frequencies. The number of peaks is found to evolve from one up to four peaks at the end of the cylinder. Such experimental measurement of increasing peaks number has not been discussed before in the literature for hypersonic boundary layers and these different peaks were previously discussed as to originate from either the first-mode, the second-mode or their non-linear interactions. A careful global linear stability analysis allows us to demonstrate that the increasing number of wall pressure peaks in the experimental spectrum is actually linear in its nature. These signatures are shown to be generated by acoustic trapped waves under the sonic line in the separated region. Each peak corresponding to a multiple of the fundamental frequency and a link between these trapped acoustic waves and the higher Mack modes is suggested. These analyses also suggest that the low frequency peaks observed in the separated region previously identified as being first-mode waves signatures may be related to low order acoustic waves, themselves related to a second mode continuation instead. Additionally, Resolvent optimal responses and SPOD modes at the leading peaks frequencies are found to be in good agreement, confirming the nature of the experimentally observed waves. Finally, phase coupling analysis at the wall using the PCB sensors bispectra highlight possible triad relations between the linearly generated acoustic waves. Such triad interactions are discussed as a mechanism of energy transfer from the higher to the lower frequencies in the separated region. However, and as highlighted on figure~\ref{fig:scenario_summary} with a dashed arrow, a complete transition scenario for this large separation case remains to be further studied. Especially, possible links with the previously studied oblique breakdown has to be investigated.

Collectively, these findings advance the understanding of transitional separated flows dynamics in the hypersonic regime. Most importantly, our results demonstrate that different transition pathways, each governed by distinct physical mechanisms, can drive the evolution of the separated region as the Reynolds number varies. These results contribute to the fundamental understanding of hypersonic boundary-layer instability and provide a basis for delineating the key mechanisms of transition for hypersonic vehicles. While this work clarifies key aspects of linear and nonlinear transition stages, further investigation of the complete transition process—particularly for large-separation cases—would benefit from additional experimental and numerical analysis. We hope that the experimental data provided here can be used as a reference to define future detailed numerical studies in order to investigate in detail the linear growth in the bubble and the discussed non-linear coupling.

\section*{Competing interests}
The authors declare no conflict of interests.

\section*{Funding statement}
This work was supported by funding from CEA and ONERA.

\newpage
\appendix

\section{Pressure profiles at $m=10$}\label{annex:prof_m10}
\begin{figure}
    \centering
    \includegraphics[width=0.9\textwidth]{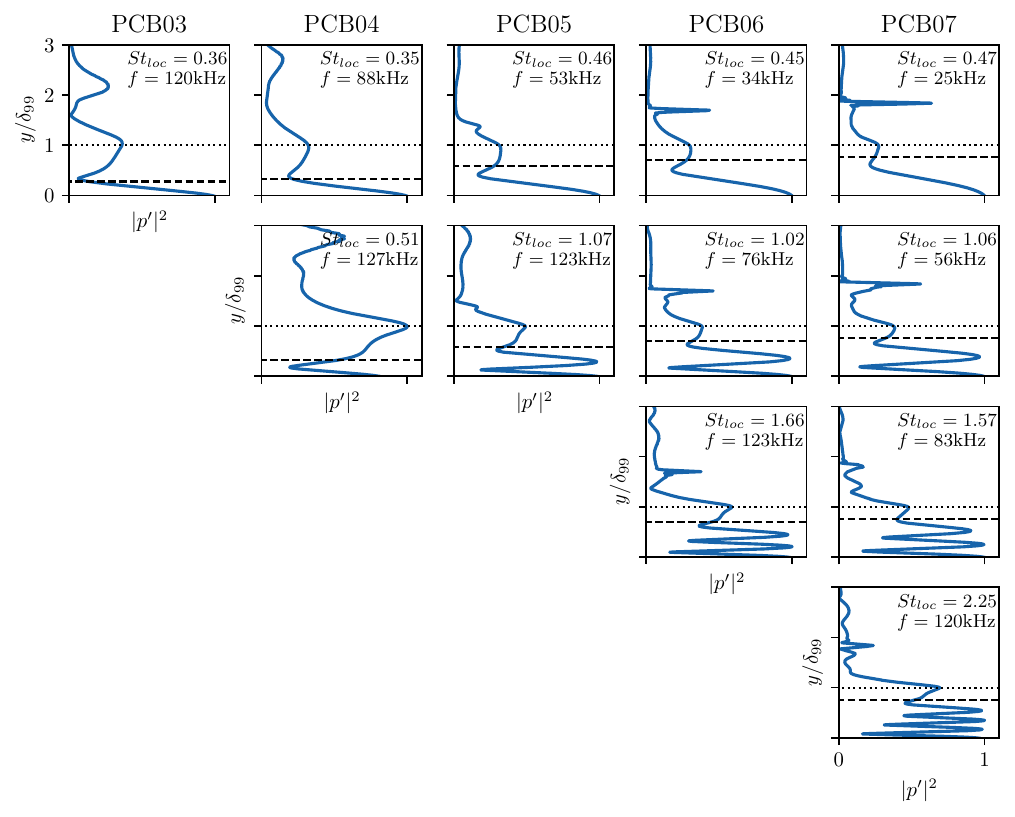}
    \caption{Streamwise evolution of the Resolvent optimal Response along the cylinder for highlighted peaks at $m=10$. 
    }
    \label{fig:case1_res10_optimal_p_prof}
\end{figure}

Equivalent results to those presented in figure \ref{fig:case1_res_optimal_p_prof}, but for a non-zero azimuthal wavenumber of 10, are shown in figure \ref{fig:case1_res10_optimal_p_prof}. They exhibit the same behaviour as the axi-symmetric structures, which supports the conclusions of the article.

\bibliographystyle{jfm}
\bibliography{jfm}

\end{document}